\documentclass[twocolumn,aps,showpacs,pre,amsmath,amssymb,floatfix,longbibliography]{revtex4-2}

\usepackage{color}
\usepackage{multirow}
\usepackage{bm}
\usepackage{epsfig}
\usepackage{psfrag}
\usepackage[ansinew]{inputenc}
\usepackage[english]{babel} 
\usepackage{amsfonts}
\usepackage{amsmath}
\usepackage{upgreek}
\usepackage{enumitem}
\usepackage[dvipsnames]{xcolor}
\usepackage{environ}  
\usepackage{hyperref}
\usepackage[extra]{tipa}
\usepackage[bb=libus]{mathalpha}

\begin{document}

\title{
Continuous Specialization Transition in the Soft Committee Machine with ReLU Activation 
	}

\author{
Assem Afanah}
\affiliation{
	Institut f\"{u}r Theoretische Physik, Universit\"{a}t
  Leipzig,  Br\"{u}derstrasse 16, 04103 Leipzig, Germany
	}

\author{
Bernd Rosenow}
\affiliation{
	Institut f\"{u}r Theoretische Physik, Universit\"{a}t
  Leipzig,  Br\"{u}derstrasse 16, 04103 Leipzig, Germany
	}	
\date{\today}

\begin{abstract}
We analyze the soft committee machine with Rectified Linear Unit (ReLU) activation by means of the replica method. In a realizable teacher--student setting, we compute the quenched free energy within a replica-symmetric ansatz and obtain the typical generalization behavior from the saddle-point equations for the macroscopic order parameters. The system exhibits a transition from an unspecialized symmetric phase to a specialized phase in which the permutation symmetry among hidden units is broken. We determine the critical training-set size as a function of the inverse training temperature and derive analytic expressions both near the transition and in the asymptotic large-sample regime. Unlike the corresponding model with sigmoidal activations, which undergoes a first-order transition, the ReLU soft committee machine shows a continuous specialization transition. These results show that the activation function plays a decisive role in the phase structure and generalization behavior of multilayer networks.
\end{abstract}
 
\maketitle
\section{Introduction}
\begin{figure}[t]
	\centering
	\includegraphics[width=6 cm, height=6 cm,keepaspectratio]{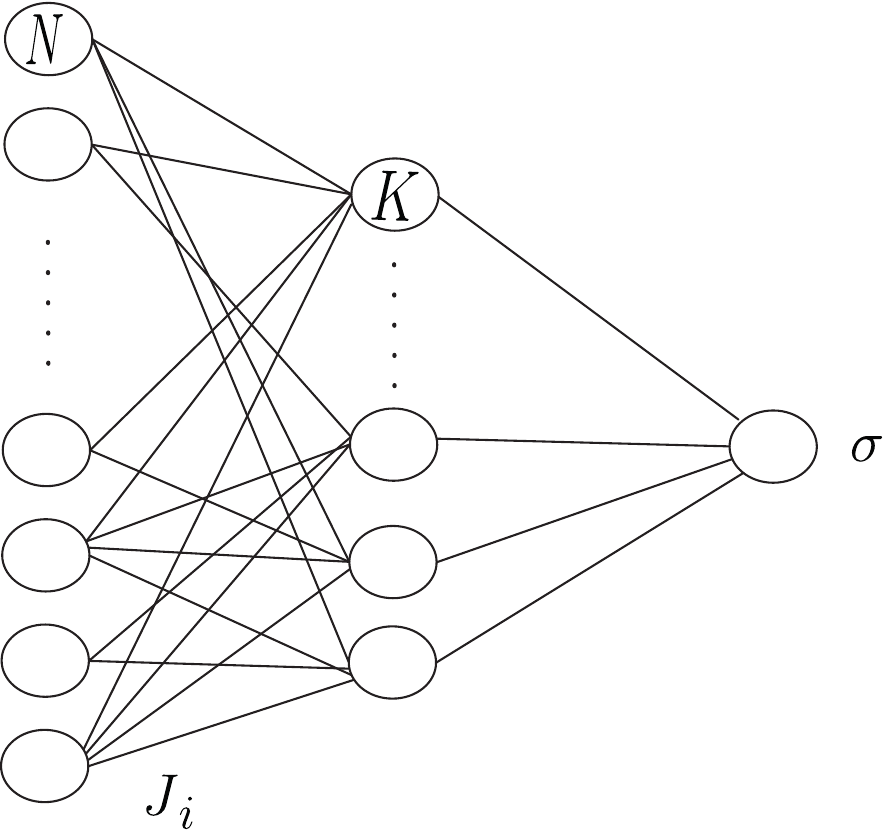}
	\caption{\label{Fig:SCM}
Schematic diagrams of the student/teacher soft committee machines. Both networks have an $N$-dimensional input layer with $K$ hidden units, we denote the student weight vectors from the input to hidden layer by $\bm{J}_{i}$ while the teacher weight vectors are denoted by $\bm{B}_{j}$; the weights from the hidden layer to the output unit are fixed to one. For a given input $\bm{\xi} \in \mathbb{R}^{N}$, the output of the SCM is proportional to the sum of the hidden-layer activations under a Rectified Linear Unit (ReLU) activation function, $g(x) = x \Theta(x)$, where $\Theta(x)$ is the Heaviside step function.  }
\end{figure}

Neural networks (NNs) have attracted significant attention over the past decade due to their remarkable success across diverse domains of science and engineering \cite{Deepbook,COLLINS2021,niskanen2023}, and they are now standard tools for supervised learning \cite{lecun2015,goodfellow2016}. 
For theoretical purposes, however, the central question is usually not the performance of a particular trained network, but the typical behavior of a large class of networks learning from random examples. 
This is naturally a problem in statistical mechanics. 
In a teacher--student setting, the training set plays the role of quenched disorder, the cost function defines an effective energy, and the Gibbs distribution over weights permits a description of learning and generalization in terms of a small number of macroscopic overlaps \cite{engel,book_shallow,nishimori2001,Watkin,Advani,phase_Bahri,Zdeborova2016InferenceReview}.  In this framework one can ask when a network begins to correlate with a target rule, when hidden units become distinguishable, and how these changes depend on the amount of data and on the training temperature.

This point of view has proved useful in a broad range of learning problems. For perceptrons and other simple models it yields typical learning curves, storage capacities, and phase transitions in a form that can be analyzed explicitly \cite{seung1992,opper1996,Watkin}. 
For multilayer networks it provides a controlled setting in which hidden-unit symmetry breaking, metastability, and specialization can be studied quantitatively \cite{H.Schwarze_1992,H.Schwarze_1993,HSchwarze2_1993,HSchwarze_1993,Ahr_1999}. More recent work has extended this program to wider teacher--student networks, where equilibrium calculations can be compared with computational thresholds, Gaussian-equivalence arguments, and the dynamics of stochastic gradient descent \cite{Aubin2019CommitteeMachine,Goldt2020DynamicsSGDTeacherStudent,Goldt2022GaussianEquivalence}. 
In that broader landscape, analytically tractable committee machines remain useful because they are simple enough to solve and yet rich enough to exhibit nontrivial collective behavior.

In this paper we study the soft committee machine (SCM), a two-layer network whose output is the average of its hidden-unit activations; see Fig.~(\ref{Fig:SCM}). The analysis is carried out in a realizable teacher--student setting, in which the student network attempts to imitate a teacher with the same architecture. The SCM is a standard model for specialization. In the unspecialized phase, the student hidden units are statistically equivalent and each unit carries only averaged information about the teacher. In the specialized phase, this permutation symmetry is broken, and different student units develop distinct overlaps with different teacher units \cite{H.Schwarze_1992,H.Schwarze_1993,HSchwarze2_1993,Ahr_1999}. The transition between these two regimes is the central collective phenomenon in the model.

We focus on the Rectified Linear Unit (ReLU),
\[
g(x)=x\Theta(x).
\]
ReLU was introduced as a simple non-saturating activation and is now standard in applications \cite{nair2010,glorot2011,relu_speech,relu_empirical}. For the present problem, however, the main point is not its empirical popularity but its geometry. ReLU is piecewise linear, non-saturating for positive arguments, and identically zero for negative arguments. These features alter the local-field statistics and, through them, the effective free energy. As a result, the activation function can influence not only quantitative learning curves but also the order of the specialization transition itself. This has already been seen in annealed studies of the ReLU committee machine and in more recent analyses of shallow networks with general activation functions \cite{Oostwal,nishiyama2024,Otavio,Afanah2025UnifiedSCM}. In particular, the contrast with the sigmoidal SCM is sharp: for sigmoidal activations the specialization transition is first order and accompanied by pronounced metastability, whereas the ReLU case appears to allow a continuous onset of specialization \cite{Ahr_1999,Oostwal}.

The purpose of the present work is to examine this question at the level of the quenched free energy. We consider the limit $N\to\infty$, $K\to\infty$ with $K/N\to0$, and compute the quenched free energy by the replica method. The calculation is carried out within a replica-symmetric and site-symmetric ansatz for the order parameters. This reduces the problem to a small set of overlaps between student and teacher weight vectors. These overlaps distinguish the unspecialized and specialized phases and determine the generalization error as a function of the scaled training-set size $\alpha=P/(NK)$ and the inverse training temperature $\beta$. The resulting theory is an equilibrium description of specialization. It does not attempt to describe the detailed out-of-equilibrium training dynamics, nor does it settle the separate question of replica-symmetry stability \cite{Engel2,Goldt2020DynamicsSGDTeacherStudent}.

Within this framework we find an unspecialized symmetric phase and a specialized phase separated by a continuous transition at a critical value $\alpha_c(\beta)$. Near the transition, the specialization order parameter scales as $(\alpha-\alpha_c)^{1/2}$. The critical training-set size decreases as $\beta$ increases and approaches a finite zero-temperature limit, $\alpha_c \approx 0.57$, as $\beta\to\infty$. In the opposite limit of high temperature, the quenched free energy reduces to the annealed result, which provides a useful check on the calculation. In the asymptotic regime of large $\alpha$, the generalization error decays as
\[
\varepsilon_g = \frac{1}{2\alpha\beta}.
\]
Thus the quenched ReLU SCM differs qualitatively from its sigmoidal counterpart, while remaining consistent with the earlier annealed description in the appropriate limit \cite{Ahr_1999,Oostwal,Afanah2025UnifiedSCM}. In Sec.~II we define the model and derive the quenched free energy. In Sec.~III we analyze the saddle-point solutions, discuss the specialization transition and its limiting forms, and obtain the asymptotic behavior of the generalization error.

\section{Method}

We use the replica method to compute the quenched free energy of the soft committee machine. The method was developed for disordered systems such as spin glasses \cite{spinNN,Mezard1987}, and has long been used in the statistical theory of neural networks and related optimization problems \cite{Hidetoshi,Advani,talagrand}. We begin with the teacher--student model. For $K=M$, the outputs of student replica $a$ and of the teacher, for an input vector $\bm{\xi}^{\mu} \in \mathbb{R}^{N}$, are
\small 
\begin{align}
\sigma^{a} = \dfrac{1}{\sqrt{K}} \sum_{i=1}^{K} g\left( \dfrac{1}{\sqrt{N}}\bm{J}^{a}_{i} \cdot \bm{\xi}^{\mu} \right) ,~ \tau = \dfrac{1}{\sqrt{K}} \sum_{j=1}^{K} g\left( \dfrac{1}{\sqrt{N}}\bm{B}_{j} \cdot \bm{\xi}^{\mu}\right).
\end{align}
\normalsize
%
Here $a=1,2,\dots,n$ labels the $n$ replicas and $g(x)$ is the ReLU activation function. The adaptive student weight vectors satisfy $(\bm{J}_{i}^{a})^{2}=N$, while the teacher vectors are mutually orthogonal,  $\bm{B}_{i}\cdot \bm{B}_{j} = N\,\delta_{ij}$. 
The training set is
$\mathbb{D}=\left\lbrace \bm{\xi}^{\mu}, \tau(\bm{\xi}^{\mu}), \mu = 1, ... , P \right\rbrace $,  
where the inputs are independent and identically distributed with unit variance in each component.
For replica $a$, the training error is measured via a quadratic cost function 
\begin{align}
\epsilon_{t} = \dfrac{1}{P} \sum_{\mu = 1}^{P} \frac{1}{2} \left[ \sigma^{a}(\bm{\xi}^{\mu}) - \tau(\bm{\xi}^{\mu})\right]^{2}\ .
\end{align}
The corresponding generalization error, i.e. the expected error on a fresh random input, is
\begin{align}
\varepsilon_{g} = \dfrac{1}{2} \left\langle \left[ \dfrac{1}{\sqrt{K}} \sum_{i=1}^{K} g(x^{a}_{i}) - \dfrac{1}{\sqrt{K}} \sum_{j=1}^{K} g(y_{j})\right]^{2}  \right\rangle_{\bm{\xi}} \ ,  \label{Eq:EgAvg_rep} 
\end{align}
where the average is over a new input $\bm{\xi}$, and the local fields are  $x^{a}_{i} = \bm{J}^{a}_{i} \cdot \bm{\xi} / \sqrt{N}$ and $y_{j} = \bm{B}_{j} \cdot \bm{\xi} / \sqrt{N}$. 
In the limit $N\to\infty$, the local fields are jointly Gaussian. The average in Eq.~(\ref{Eq:EgAvg_rep}) can therefore be expressed in terms of the macroscopic overlaps 
 $Q^{aa}_{ij} = \bm{J}^{a}_{i} \cdot \bm{J}^{a}_{j} / N $ and $R_{ij} = \bm{J}^{a}_{i} \cdot \bm{B}_{j} / N $, which are self-averaging in the thermodynamic limit. One obtains \cite{Oostwal}
\begin{align}
\nonumber \varepsilon_{g}^{a} =  \, & \dfrac{1}{2 K} \, \sum_{i,j=1}^{K} \left( \dfrac{Q^{aa}_{ij}}{4} + \dfrac{\sqrt{1- (Q^{aa}_{ij})^{2}}}{2 \pi} + \dfrac{Q^{aa}_{ij} \arcsin [Q^{aa}_{ij}]}{2 \pi} \right) \\
\nonumber & - \dfrac{1}{K} \sum_{i,j=1}^{K} \left( \dfrac{R^{a}_{ij}}{4} + \dfrac{\sqrt{1- (R^{aa}_{ij})^{2}}}{2 \pi} + \dfrac{R^{a}_{ij} \arcsin [R^{a}_{ij}]}{2 \pi} \right) \\
& + \left( \dfrac{1}{2} +  \dfrac{K - 1}{4 \pi} \right).   \label{Eq:Eg_replica1}  
\end{align} 

Following Ahr et al.~\cite{Ahr_1999}, we evaluate the disorder average of $\ln Z$, and hence the quenched free energy, by means of the replica identity
\begin{align}
\left\langle \text{ln} Z \right\rangle =   \dfrac{\partial \left\langle Z^{n} \right\rangle}{\partial n}\Bigg|_{n=0}  \ .
\label{Eq:repZ} 
\end{align}
Here $Z$ is the Gibbs partition function of a single system, and $Z^{n}$ is the partition function of $n$ noninteracting replicas. Averaging over the independent training examples gives
\begin{align}
\left\langle Z^{n} \right\rangle =  \int \prod_{a=1}^{n} \prod_{i=1}^{K}  d\mu(\bm{J}_{i}^{a})  \text{exp}(-P G_{e})  \, \label{Eq:Zn} 
\end{align}
where $d\mu(\bm{J}_{i}^{a})$ denotes the measure enforcing $(\bm{J}_{i}^{a})^{2}=N$, and
\begin{align}
G_{e} = - \text{ln} \left\langle \text{exp}\left[\dfrac{-\beta}{2} \sum_{a=1}^{n} \left[ \sigma^{a}(x^{a}) -  \tau(y)\right]^{2}  \right] \right\rangle_{\xi} \label{Eq:Ge} 
\end{align}
is the energetic contribution. To evaluate $G_{e}$, we introduce the vector  $\bm{\sigma} = \left( \sigma^{1}, \sigma^{2}, .....,\sigma^{n}, \tau \right)^{T} $ so that 
 $\sum_{a=1}^{n} (\sigma^{a}- \tau)^{2} = \bm{\sigma}^{T} \Sigma \bm{\sigma}$ with the $(n+1)\times (n+1)$ matrix 
\begin{align}
\Sigma = \begin{pmatrix}
1 &0&.....&-1 \\
0&1&.....& -1 \\
\vdots &\vdots&\ddots&\vdots \\
-1&-1& \cdots &n
\end{pmatrix} .
\end{align}
In the large-$K$ limit, $\bm{\sigma}$ is Gaussian with mean 
\begin{align}
\bm{\mu} = \left( <\sigma^{1}>, <\sigma^{2}>, .....,<\sigma^{n}>, <\tau> \right)^{T} .
\end{align}
It is therefore convenient to define the centered variables $\tilde{\sigma}^{a} = \sigma^{a} - <\sigma^{a}>$, $\tilde{\tau} = \tau - \langle \tau \rangle$, and   
$\bm{\tilde{\sigma}} = \left( \tilde{\sigma}^{1}, \tilde{\sigma}^{2}, .....,\tilde{\sigma}^{n}, \tilde{\tau} \right)^{T} $.  
The joint distribution of $\bm{\tilde{\sigma}}$ is
\begin{align}
    P(\bm{\tilde{\sigma}}) = \dfrac{1}{\sqrt{(2\pi)^{n+1} \vert M \vert}} \, \text{exp}\left[-\dfrac{1}{2} \bm{\tilde{\sigma}}^{T} M^{-1} \bm{\tilde{\sigma}}  		
    \right] \ ,
\end{align} 
which is completely specified by the covariance matrix $ M= \left\langle \bm{\tilde{\sigma}} \, \bm{\tilde{\sigma}}^{T} \right\rangle $. Using this notation, the average in $G_{e}$ is now an elementary Gaussian integral \cite{Ahr_1999}  :
\small 
\begin{align}
\nonumber     \left\langle \text{exp}\left[ -\dfrac{\beta}{2} 
		       \bm{\tilde{\sigma}}^{T} \Sigma \bm{\tilde{\sigma}} \right] 
		       \right\rangle  
        	   & = \\
\nonumber      \dfrac{(2\pi)^{- (n+1)/2}}{\sqrt{\vert M \vert}} \, & \int 
		        d\bm{\tilde{\sigma}}^{n+1} \text{exp}\left[ -\dfrac{1}{2} 
		       \bm{\tilde{\sigma}}^{T} (\beta \Sigma+  M^{-1}) \bm{\tilde{\sigma}} \right] \\
		       &= \dfrac{1}{\sqrt{\vert\beta M \Sigma + I\vert}} \, .	                                            
\end{align}
\normalsize
Thus, we obtain the energetic contribution
\begin{align}
 G_{e} = \dfrac{1}{2} \text{ln} \left[ \det (\beta M \Sigma + I)\right]. \label{Eq:Gr_eff}   
\end{align}

To expose the dependence on the macroscopic overlaps, we introduce the order parameters $(Q^{ab}_{ij},R^{a}_{ij})$ into Eq.~(\ref{Eq:Zn}) by means of delta functions. This generates an entropic contribution $G_{s}$ and leads to
\begin{align}
	\left\langle Z^{n}\right\rangle =  \int \prod_{a,b=1}^{n} \, \prod_{i,j=1}^{K}  dQ_{ij}^{ab} \, dR_{ij}^{a} \, \text{exp}(-P G_{e}+ N G_{s})\ . \label{Eq:Zn2} 	
\end{align}
If the number of examples scales as $P=\alpha N K$, this integral is dominated by a saddle point in the limit $N\to\infty$. The entropic term is
\begin{align}
\nonumber G_{s} = \dfrac{1}{N} \, \text{ln} \int & \prod_{a,b=1}^{n} \prod_{i,j=1}^{K} \, d\mu(\bm{J}_{i}^{a}) \, \delta(N Q_{ij}^{ab}- \bm{J}^{a}_{i}
	 \cdot \bm{J}^{b}_{j} ) \\
	 & \times \delta(N R_{ij}^{a}- \bm{J}^{a}_{i}\cdot \bm{B}_{j} )\ .
\end{align} 
Using the integral representation of the delta functions and evaluating the resulting integrals by saddle point, one obtains \cite{Ahr_1999}
\begin{align}
	G_{s} = \dfrac{1}{2} \, \text{ln}(\det \mathcal{C}) + \text{const.} \ . \label{Eq:Gs}
\end{align}
where $\mathcal{C}$ is the $[(n+1)K]\times[(n+1)K]$ matrix of all student--student, student--teacher, and teacher--teacher overlaps,
\begin{align}
\mathcal{C} = \begin{pmatrix}
		Q^{nK \times nK}  & R^{nK \times K} \\
		R^{T}  & T^{K \times K}
		 \end{pmatrix} \; .
\end{align} 
Because the teacher vectors are orthonormal, the teacher--teacher block is simply the $K\times K$ identity matrix.  To simplify the energetic and entropic terms, we consider the limit $K\to\infty$ with $K/N\to0$ and adopt a site-symmetric, replica-symmetric ansatz,
\begin{align}
\nonumber	Q_{ij}^{aa} = \begin{cases}
				  1 & \text{if} \, i=j\\
				  C & \text{if} \, i\neq j
				  \end{cases} ~ ,& \qquad
	Q_{ij}^{ab}	= \begin{cases}
				  q & \text{if} \, i=j\\
				  p & \text{if} \, i\neq j
				  \end{cases} \quad a \neq b\\
	R_{ij}^{a}  &= \begin{cases}
				  R & \text{if} \, i=j\\
				  S & \text{if} \, i\neq j
				  \end{cases}\ .  \label{Eq:ansat_rep}
\end{align} 
As in Ref.~\cite{Ahr_1999}, we further assume that $(C,p,S)$ are of order $1/K$, and therefore write
 $S= \hat{S}/K,~ C= \hat{C}/(K-1)$ and $ p=\hat{p}/K$. 
To characterize specialization of the hidden units, we define $\Delta = R-S$ and $\delta = q-p$. The remaining step is to evaluate the determinants in Eqs.~(\ref{Eq:Gr_eff}) and (\ref{Eq:Gs}) and then perform the analytic continuation $n\to0$; details are given in Appendix~A. This yields the free-energy density
\small
\begin{equation}
\begin{split}
f \equiv \dfrac{2\beta F}{N K} & = \alpha \left[ \dfrac{\beta (v-2w + (1/2 - 1/2\pi))}{1+\beta (u-v)} + \text{ln}[1+\beta (u-v)] \right] \\
                                & + \dfrac{\delta - \Delta^{2}}{\delta - 1} - \text{ln}(1-\delta) -  \dfrac{\delta + \hat{p} - (\Delta + \hat{S})^{2}}{\tilde{C}} + \mathcal{O}(\dfrac{1}{K})
\end{split} \label{Eq:free_rep} 
\end{equation} 
\normalsize
where 
\begin{subequations}
\begin{align}
\tilde{C} &= K(1+ \hat{C} - \delta - \hat{p}) \\
u &= \dfrac{\hat{C}}{4} + \left(\dfrac{1}{2}- \dfrac{1}{2\pi} \right) \\
v &= \dfrac{\delta}{4} + \dfrac{\hat{p}}{4} + \dfrac{\sqrt{1- \delta^{2}}}{2 \pi} + \dfrac{\delta \arcsin[\delta]}{2\pi} - \dfrac{1}{2\pi} \\
w &= \dfrac{\Delta}{4} + \dfrac{\hat{S}}{4} + \dfrac{\sqrt{1- \Delta^{2}}}{2 \pi} + \dfrac{\Delta \arcsin[\Delta]}{2\pi} - \dfrac{1}{2\pi} .  
\end{align}
\label{Eq:uvw}
\end{subequations} 
Because of the scaling introduced above, the free energy is expressed in terms of variables of order unity. This is the form used below for both analytic expansions and numerical solution of the saddle-point equations in the symmetric, specialized, and asymptotic regimes. 
Finally, Eq.~(\ref{Eq:Eg_replica1}) becomes
\begin{align}
\varepsilon_{g} = \dfrac{\hat{C}}{8} - \left( \dfrac{\Delta}{4} + \dfrac{\hat{S}}{4} + \dfrac{\sqrt{1- \Delta^{2}}}{2 \pi} + \dfrac{\Delta \arcsin[\Delta]}{2\pi} \right) + \dfrac{1}{2}\ .    
\end{align}

\section{Results and Discussion}
\begin{figure*}[t!]
	\centering
	\includegraphics[width=12 cm, height=12 cm,keepaspectratio]{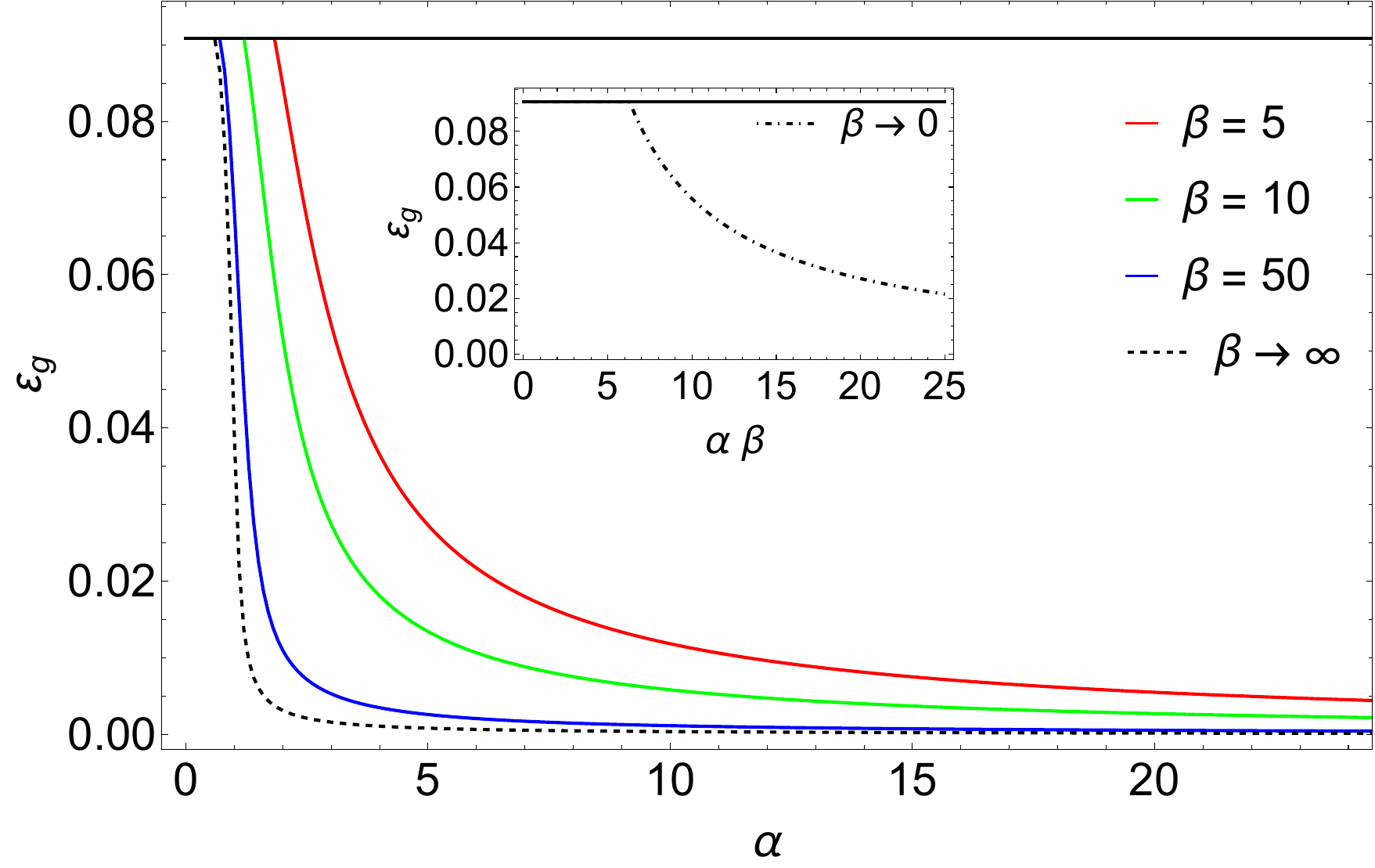}
	\caption{\label{Fig:rep_relu} 
    Generalization error as a function of $\alpha$ for several values of the inverse training temperature $\beta$. For each $\beta$, the system undergoes a continuous transition at $\alpha_{c}(\beta)$ from an unspecialized symmetric phase with $(\Delta=\delta=0)$, which gives a plateau at $\varepsilon_{g}=1/4-1/2\pi$, to a specialized phase with $(\Delta,\delta>0)$.
    The plateau becomes shorter as $\beta$ increases. The dashed curve shows the limit $\beta\rightarrow\infty$, for which the critical value approaches $\alpha_{c}\approx0.57$. The inset shows the high-temperature limit $\beta\rightarrow0$, plotted as a function of the scaled variable $\alpha\beta$. In this limit the transition occurs at $(\alpha\beta)_{c}\approx2\pi$, in agreement with the annealed approximation.}
\end{figure*}

The physical solutions are obtained from the saddle-point equations of the free energy. As in Ref.~\cite{Ahr_1999}, the condition $\partial f/\partial \hat{S}=0$ implies that $\tilde{C}$ must remain of order $\mathcal{O}(1)$. The saddle-point equations then give (see Appendix~B)
\begin{subequations}
\label{Eq:sol_21}
\begin{align}
\hat{p} =& 1- \delta\ , \label{Eq:sol_1}\\
\hat{S} =& 1- \Delta\ , \label{Eq:sol_2}\\
\hat{C} =& 0 \ . \label{Eq:sol_3}
\end{align}
\end{subequations}
The remaining order parameters, $\delta$ and $\Delta$, must in general be determined numerically as functions of $\alpha$ and $\beta$. Their behavior simplifies, however, both near the transition and in the asymptotic large-$\alpha$ regime, where analytic expansions are possible.
Figure~(\ref{Fig:rep_relu}) shows the generalization error for several values of $\beta$. There are two branches of solutions, the first is the unspecialized symmetric solution,
\begin{subequations}
\begin{align}
\Delta =& \delta = 0 \\
\hat{p} =& \hat{S} = 1 \ ,
\end{align}
\end{subequations}
for which
\begin{align}
\varepsilon_{g} = \dfrac{1}{4} - \dfrac{1}{2\pi}
\ \end{align}
independent of $\alpha$ and $\beta$.
The second is a specialized solution with $(\Delta,\delta)>0$, which appears above a critical value $\alpha_{c}(\beta)$. The transition corresponds to the breaking of the permutation symmetry among the student hidden units.

\begin{figure*}[t!]
	\centering
	\includegraphics[width=15 cm, height=10 cm,keepaspectratio]{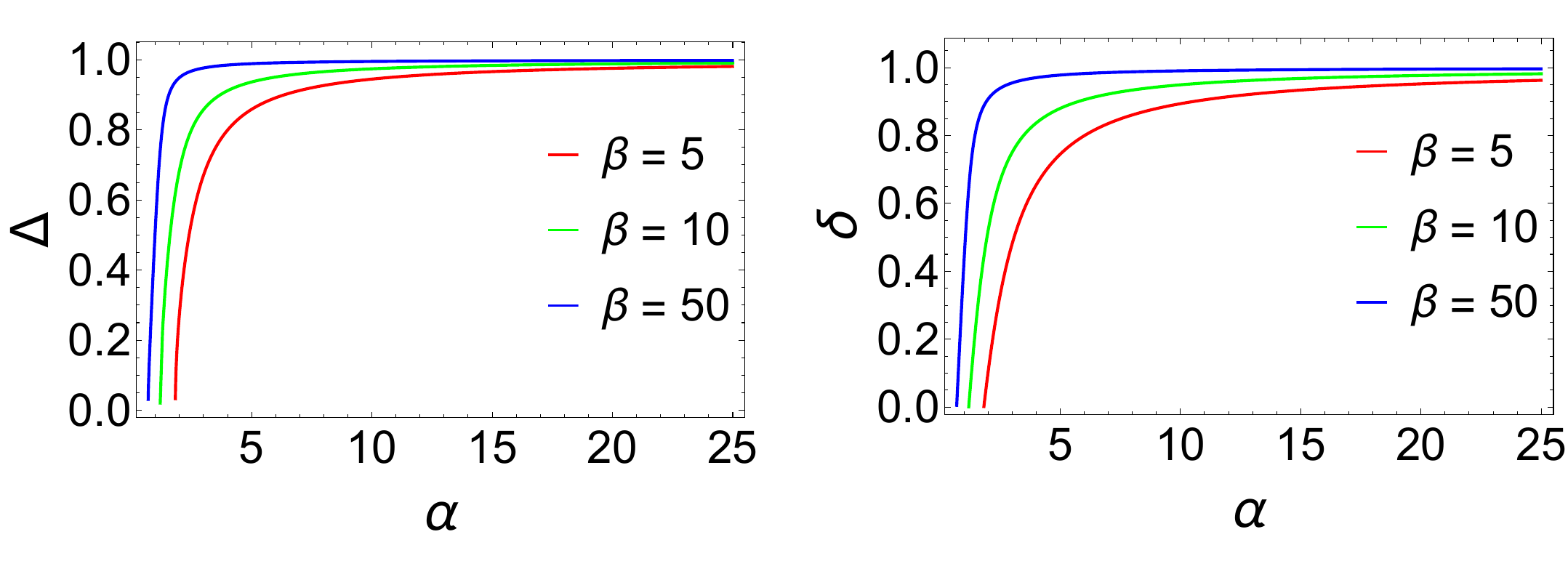}
	\caption{\label{Fig:Delta_rep} 
   Order parameters $\Delta=R-S$ and $\delta=q-p$ as functions of $\alpha$ for $\alpha>\alpha_{c}$. Both increase monotonically with $\alpha$ and therefore measure the degree of specialization in the network. Asymptotically, $\Delta,\delta\rightarrow1$, corresponding to perfect alignment between student and teacher and hence $\varepsilon_{g}\rightarrow0$. 
    }
\end{figure*}

As $\alpha$ increases beyond $\alpha_{c}$, the specialization becomes stronger and both order parameters approach unity, $(\Delta,\delta)\rightarrow1$, as shown in Fig.~(\ref{Fig:Delta_rep}).  In the present realizable setting with $K=M$, this corresponds to one-to-one alignment of the student hidden units with the teacher hidden units, up to permutation. In replica language, all replicas select the same representative of the version space \cite{Ahr_1999}. Consequently, the generalization error tends to zero in the asymptotic regime.
 The dependence on $\beta$ is shown clearly in Fig.~(\ref{Fig:rep_relu}). As $\beta$ increases, the unspecialized plateau becomes shorter and specialization sets in at smaller $\alpha$. The two limiting cases, $\beta\rightarrow0$ and $\beta\rightarrow\infty$, show the same overall structure and will be discussed in more detail in Sec.~\ref{sec:large_low_beta}.

\subsection{Solutions in the vicinity of $\alpha_{c}$}
\begin{figure*}[t!]
	\centering
	\includegraphics[width=15 cm, height=15 cm,keepaspectratio]{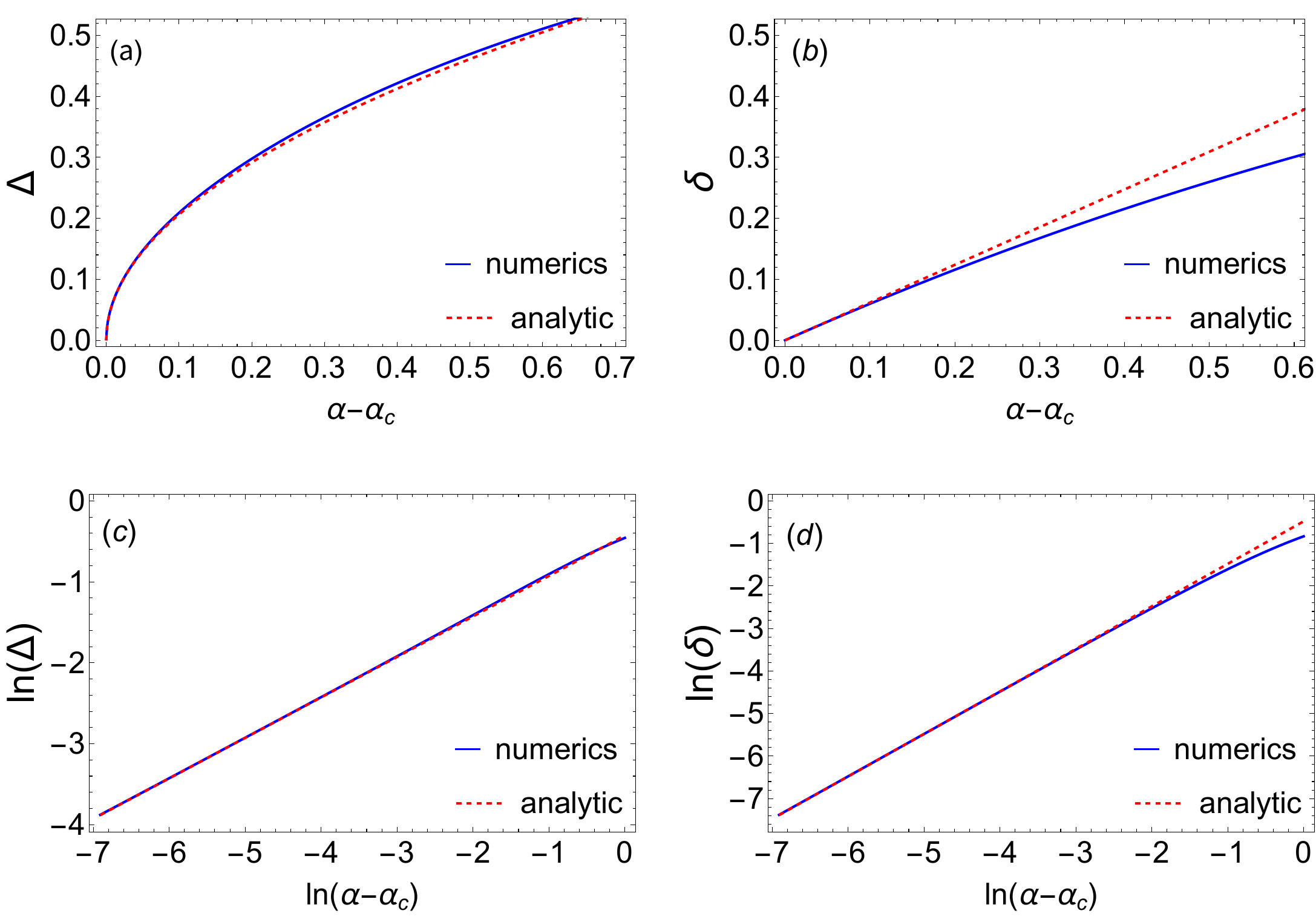}
	\caption{\label{Fig:analytVsnum} 
    Order parameters $(\Delta,\delta)$ close to the transition point $\alpha_{c}$ for $\beta=5$. In panels (a) and (b), the analytic results (red dashed curves) agree closely with the numerical solutions (solid blue curves) near $\alpha_{c}$ and deviate only farther away from the transition. The log-log plots in panels (c) and (d) show the expected scaling, $\Delta\propto(\alpha-\alpha_{c})^{1/2}$ and $\delta\propto(\alpha-\alpha_{c})$, consistent with Eq.~(\ref{Eq:sol_anlyt}).
     }
\end{figure*}

To analyze the onset of specialization, we insert $\hat{p}=1-\delta$ and $\hat{S}=1-\Delta$ into the free energy and expand for small $(\Delta,\delta)$. Since specialization is absent at quadratic order alone, it is necessary to retain terms up to $\mathcal{O}(\Delta^{4},\delta^{2})$. This gives
\begin{align}
f = const. - c_{1} \delta^{2} + c_{2} \Delta^{2} - c_{3} \Delta^{4} + \Delta^{2} \delta + \mathcal{O}(\Delta^{5},\delta^{3}) \ , \label{Eq:free_rep_exp}
\end{align}
where  
\begin{align}
\nonumber  c_{1} = -\dfrac{\alpha \tilde{\beta}^{2}}{8 \pi} \left({1 \over 2} - \dfrac{1}{ \pi}\right) + \dfrac{1}{2} \, , & \quad 
           c_{2} = \dfrac{2 \pi - \alpha \tilde{\beta} }{2 \pi}\ ,  \\
           c_{3} =& \dfrac{\alpha \tilde{\beta}}{24 \pi} \ ,
\end{align}
with $\tilde{\beta} = \dfrac{\beta}{1+ \beta (1/4 - 1/2\pi)}$. 
For fixed $\beta$, $c_{1}$ remains positive in the regime of interest, while $c_{2}$ changes sign at the transition. The condition $c_{2}=0$ gives 
\begin{align}
 \alpha_{c} = \dfrac{2\pi}{\beta} + \dfrac{\pi -2}{2}\ .
\label{Eq:alphac} 
\end{align} 
The corresponding saddle-point equations are
\begin{subequations}
\begin{align}
-\, 2\, c_{1}\,  \delta\, + \, \Delta^{2} &= 0  \label{Eq:trans_1}\\
2\, c_{2}\,  \delta \, -\,  4 \, c_{3}\, \Delta^{3} \, +\,  2\, \delta\, \Delta &= 0 \ .
\label{Eq:trans_2}
\end{align}   
\end{subequations}
From the first equation,
\begin{align}
\delta =  \dfrac{\Delta^{2}}{2 c_{1}}  \ .
\label{Eq:trans_sol_1}
\end{align}
Substituting into Eq.~(\ref{Eq:trans_2}) gives
\begin{align}
\Delta^{2} = \dfrac{2 c_{1} c_{2}}{4 c_{1} c_{3}-1} \ . 
\label{Eq:trans_sol_2}
\end{align} 
For $\alpha<\alpha_{c}$, the only real solution is the symmetric one, $\Delta=\delta=0$. For $\alpha>\alpha_{c}$, a branch with $\Delta,\delta>0$ appears continuously. It is useful to eliminate $\delta$ by means of Eq.~(\ref{Eq:trans_sol_1}). This reduces the Landau expansion to
\begin{equation}
f_{\mathrm{eff}} = \text{const.} + c_{2}\Delta^{2}
+ \left(\dfrac{1}{4c_{1}}-c_{3}\right)\Delta^{4}
+ \mathcal{O}(\Delta^{6}) \ , \label{freenergy_Delta.eq}
\end{equation}
so the onset of specialization is controlled by the sign change of $c_{2}$. The quartic term stabilizes the specialized branch, and the transition is therefore continuous.

For completeness, the determinant of the Hessian in the full $(\Delta,\delta)$ description is 
\begin{align}
\nonumber \det(H) =& -4 c_{1} c_{2} - 4 c_{1} \delta + \Delta^{2} (24 c_{1} c_{3} - 4) \\
        =& -4 c_{1} c_{2} + 6 \Delta^{2} (4 c_{1} c_{3} -1 )
 \label{hessian.eq} \ , \end{align}
 where in the second line we have substituted $\delta$ from Eq.~(\ref{Eq:trans_sol_1}). In the unspecialized regime, the contribution of the first term is negative while the second term vanishes with $\Delta =0$, i.e. the determinant of the Hessian is negative. 
 The appearance of a negative Hessian indicates a stable saddle point solution in replica calculations, and is due to the fact that in the replica limit $n\to 0$ the number of off-diagonal order paramters becomes negative. We note that the Hessian Eq. \ref{hessian.eq} refers only to the curvature within the reduced $(\Delta,\delta)$ manifold. It should not be interpreted as an Almeida--Thouless or replicon stability criterion, which would require fluctuations outside the present replica-symmetric, site-symmetric ansatz \cite{Almeida_1978,Castellani_2005}.

 Expanding Eqs.~(\ref{Eq:trans_sol_1}) and (\ref{Eq:trans_sol_2}) to first order in $(\alpha-\alpha_{c})$ gives
\begin{subequations}
\label{Eq:sol_anlyt}
\begin{align}
\delta &= \dfrac{12 \tilde{\beta} (\alpha - \alpha_{c})}{\pi(20+\tilde{\beta})- 2 \tilde{\beta}} \\
\Delta^{2} &= \dfrac{3 \tilde{\beta} (\alpha - \alpha_{c})\left[ 2 \tilde{\beta} - \pi(\tilde{\beta}-4)\right] }{\pi[(20+\tilde{\beta})\pi- 2 \tilde{\beta}]} \ .
\end{align} 
\end{subequations} 
The generalization error then becomes
\begin{align}
\varepsilon_{g} = \left( \dfrac{1}{4} - \dfrac{1}{2\pi} \right) - \dfrac{\Delta^{2}}{4\pi} + \mathcal{O}(\Delta^{3})\  ,
\end{align}
so $\varepsilon_{g}$ decreases linearly in $(\alpha-\alpha_{c})$ near the transition.
Figure~(\ref{Fig:analytVsnum}) compares these analytic expressions with the numerical solutions of the full saddle-point equations obtained from Eq.~(\ref{Eq:free_rep}) for $\beta=5$. Panels (a) and (b) show very good agreement close to $\alpha_{c}$, while deviations appear farther from the transition, where the truncated expansion in Eq.~(\ref{Eq:free_rep_exp}) is no longer quantitatively accurate. 
The log-log plots in panels (c) and (d) confirm the scaling laws $\Delta\propto(\alpha-\alpha_{c})^{1/2}$ and $\delta\propto(\alpha-\alpha_{c})$.

\subsection{Solutions in the asymptotic regime $\alpha \rightarrow \infty $}

We next consider the asymptotic regime $\alpha\rightarrow\infty$, where $(\Delta,\delta)\rightarrow(1,1)$. We therefore write  $\Delta = 1- \tilde{\Delta}$ and $\delta = 1- \tilde{\delta}$ with $\tilde{\Delta},\tilde{\delta}\ll1$. Using $\hat{p}=1-\delta=\tilde{\delta}$ and $\hat{S}=1-\Delta=\tilde{\Delta}$, the free energy becomes
\begin{align}
\nonumber f = & const. + \alpha \left[ \dfrac{\beta \left( \tilde{\Delta}/2 - \tilde{\delta}/4 \right) }{1+ \beta \tilde{\delta}/4} + \text{ln}\left( 1+ \beta \tilde{\delta}/4 \right) \right] \\
& - \dfrac{1- (1-\tilde{\Delta})^{2}}{\tilde{\delta}} - \text{ln}(\tilde{\delta}) + \mathcal{O}(\tilde{\Delta}^{3/2}, \tilde{\delta}^{3/2} ) \ . 
\label{Eq:free_asymp_1} 
\end{align}
Since $\tilde{\Delta}$ and $\tilde{\delta}$ vanish asymptotically, we expand the nonlinear terms and obtain
\begin{align}
f = \dfrac{\alpha \beta \tilde{\Delta}}{2} - \dfrac{2 \tilde{\Delta}}{\tilde{\delta}} - \text{ln}(\tilde{\delta}) + \mathcal{O}(\tilde{\Delta}^{2},\tilde{\delta}^{2})\ ,
\end{align}
while the generalization error depends only on $\tilde{\Delta}$ to leading order,
\begin{align}
\varepsilon_{g} = \frac{\tilde{\Delta}}{4} + \mathcal{O}(\tilde{\Delta}^{3/2})\ .
\end{align}    
The saddle-point equations now give
\begin{subequations}
\begin{align}
\tilde{\delta} =& \dfrac{4}{\alpha \beta} \\
\tilde{\Delta} =& \dfrac{2}{\alpha \beta} \ .
\end{align}
\end{subequations}
Hence the generalization error decays as
\begin{align}
\varepsilon_{g} = \dfrac{1}{2 \alpha \beta} \ .
\end{align} 
This asymptotic law is the same as that found for the SCM with error-function activation in the replica calculation of Ref.~\cite{Ahr_1999}, and it is also consistent with the annealed ReLU result reported in Ref.~\cite{Oostwal}. At large $\alpha$, the leading behavior is therefore insensitive to the choice of activation function.
The main distinction between ReLU and sigmoidal activations lies instead in the transition region and in the order of the specialization transition.

\subsection{Learning behavior in the large and low temperature limit \label{sec:large_low_beta} }

Two limiting cases are especially useful: the high-temperature limit $\beta\rightarrow0$ and the zero-temperature limit $\beta\rightarrow\infty$. The first provides a check on the quenched calculation, since the replica result must reduce to the annealed approximation in this limit. For $\beta\ll1$, we expand the energetic term of the free energy:
\small
\begin{align}
\nonumber G_{e} \approx & \alpha \left[ \beta ( v-2 w + (1/2 - 1/2\pi)) (1- \beta (u-v)) + \beta (u-v) \right]\\  
                \approx & \alpha \beta \left[( u-2 w + (1/2 - 1/2\pi)) + \mathcal{O}(\beta^{2})\right] \ .
\end{align}
\normalsize
To obtain a nontrivial limit, one keeps the product $\alpha\beta$ fixed; for notational simplicity we continue to denote this scaled variable by $\alpha$. The free energy then becomes
 
\small
\begin{align}
\nonumber f & = \alpha \left[ \dfrac{\delta}{4} + \dfrac{\hat{p}}{4} - 2 \left( \dfrac{\Delta}{4} + \dfrac{\hat{S}}{4} + \dfrac{\sqrt{1- \Delta^{2}}}{2 \pi} + \dfrac{\Delta \arcsin[\Delta]}{2\pi}  \right)  \right] \\
            & + \dfrac{\delta - \Delta^{2}}{\delta - 1} - \text{ln}(1-\delta) -  \dfrac{\delta + \hat{p} - (\Delta + \hat{S})^{2}}{\tilde{C}} + const. \label{Eq:free_b0} 
\end{align}
\normalsize
where we used the explicit forms of $u$ and $w$. In addition to Eqs.~(\ref{Eq:sol_21}), the saddle-point equations now imply $\delta=\Delta^{2}$, so the problem reduces to a single equation for $\Delta$. The same qualitative behavior is found as before: there is a continuous transition at $(\alpha\beta)_{c}\approx2\pi$, in agreement with the annealed result of Ref.~\cite{Oostwal}. This is the behavior shown in the inset of Fig.~(\ref{Fig:rep_relu}).

In the opposite limit, $\beta\rightarrow\infty$, the factor $1+\beta(u-v)$ is dominated by the term linear in $\beta$, so that $1+\beta (u-v) \approx \beta (u-v)$. The free energy becomes
\begin{equation}
\begin{split}
f &= const. + \alpha \left[ \dfrac{v-2w + (1/2 - 1/2\pi)}{u-v} + \text{ln}[u-v] \right] \\
  & + \dfrac{\delta - \Delta^{2}}{\delta - 1} - \text{ln}(1-\delta) - \dfrac{\delta + \hat{p} - (\Delta + \hat{S})^{2}}{\tilde{C}} \ .
\end{split}  \label{Eq:free_binf} 
\end{equation} 
In this zero-temperature limit the free energy, and therefore the saddle-point equations, become independent of $\beta$.  The solutions again satisfy Eqs.~(\ref{Eq:sol_21}), with $\Delta$ and $\delta$ determined numerically.  The transition occurs at $\alpha_{c}\approx0.57$, as shown by the dashed curve in Fig.~(\ref{Fig:rep_relu}). This is the lower bound approached by $\alpha_{c}(\beta)$ as $\beta$ increases.

\section{Conclusion}

In this paper we studied the soft committee machine with ReLU activation in a realizable teacher--student setting by computing the quenched free energy within a replica-symmetric, site-symmetric ansatz.
This gives an equilibrium description of generalization in terms of a small set of macroscopic overlaps and provides a simple characterization of specialization of the hidden units. 
The main result is that the ReLU soft committee machine has an unspecialized symmetric phase and a specialized phase separated by a continuous transition. This is qualitatively different from the corresponding sigmoidal model, where the specialization transition is first order and accompanied by pronounced metastability \cite{Ahr_1999}.
Within the present framework, the activation function therefore affects not only quantitative learning curves, but also the structure of the free-energy landscape and the manner in which specialization sets in \cite{nishiyama2024,Otavio}.

A second result concerns the role of the inverse training temperature $\beta$. We found that the critical training-set size is
$\alpha_{c}(\beta)=\frac{2\pi}{\beta}+\frac{\pi-2}{2}$, 
so that $\alpha_{c}$ decreases monotonically with increasing $\beta$ and approaches the finite zero-temperature limit $\alpha_{c}\approx0.57$ as $\beta\rightarrow\infty$. Thus lower training temperature favors earlier specialization.
 
We also analyzed the behavior near the transition and in the asymptotic regime. Close to $\alpha_{c}$, the order parameters obey
$ \Delta \propto (\alpha-\alpha_{c})^{1/2},
\qquad
\delta \propto (\alpha-\alpha_{c})$, and the generalization error decreases linearly in $(\alpha-\alpha_{c})$.

In the opposite limit $\alpha\rightarrow\infty$, the system approaches perfect specialization, with $\Delta,\delta\rightarrow1$, and the generalization error decays as
$\varepsilon_{g}=\frac{1}{2\alpha\beta}$.
Thus the leading large-$\alpha$ behavior agrees with earlier results for the soft committee machine with other activation functions \cite{Ahr_1999,Oostwal}. The principal distinction between ReLU and sigmoidal activations lies not in the asymptotic decay itself, but in the onset of specialization and the order of the transition.

The scope of the present analysis should also be kept in mind. Our calculation is performed within a replica-symmetric, site-symmetric equilibrium ansatz. It does not address out-of-equilibrium training trajectories, sequential specialization, or the stability of the replica-symmetric solution. These are natural directions for further work. In particular, it would be useful to examine whether replica-symmetry-breaking effects modify the transition or the structure of the specialized phase \cite{Malzahn_1999,Agliari_2020,Hartnett,Annesi2025FullRSBTwoLayer}. Another open problem is the regime $K \gtrsim N$, and especially the ultra-wide limit, where committee-machine models may help connect the statistical-mechanics description more directly to modern overparameterized networks and their improved generalization behavior \cite{doubledecent,PhysRevLett.87.078101,Barbier25,Afanah2025UnifiedSCM}. In that sense, the present work should be viewed as a controlled step toward a broader statistical-mechanical theory of specialization and generalization in multilayer networks.

\section{Acknowledgment} 
 We thank Frederieke Richert and Otavio Citton from the University of Groningen for stimulating discussions during their visit to the Institute of Theoretical Physics, Leipzig university.
       

\appendix
\section{ Derivation of the energetic and entropic terms of the free energy }  
In order to obtain the quenched free energy Eq.~(\ref{Eq:free_rep}), one need to compute 
\begin{align}
	f \doteq \dfrac{2 \beta F}{NK} = \dfrac{\partial}{\partial n} \left[ 2\alpha G_{e} - \dfrac{2}{K} G_{s} \right]_{n=0}
\end{align}
with the energetic term Eq.~(\ref{Eq:Gr_eff}) and the entropic term Eq.~(\ref{Eq:Gs}). We start with the energetic term, the matrix $\bm{M}$ takes the form
\begin{align}
	M = \begin{pmatrix}
			u &v &v &\cdots &w\\
			v &u &v &\cdots &w\\
			v &v &u &\cdots  &w\\
			\vdots &\vdots &\vdots &\ddots  &\vdots\\
			w &w &w &\cdots  & t
	\end{pmatrix} ,
\end{align}
where $u,v$ and $w$ are defined the same as in Eq.~(19) while $t = 1/2 - 1/2\pi$. For convince we write the whole matrix $(\beta M \Sigma + I)$ as 
\begin{align}
\beta M \Sigma + I = \begin{pmatrix}
							a &b &\cdots &c\\
							b &a &\cdots  &c\\
							\vdots &\vdots &\ddots  &\vdots\\	
							d &d &\cdots  &e
                               \end{pmatrix} ,\label{mat_M} 
\end{align}
with 
\begin{align*}
a = \beta(u-w)&+1, \; b = \beta(v-w), \; c = -\beta[u+ (n-1)v -nw] \\
 d &= \beta(w- t), \; e = - n \beta(w-t)+1 .
\end{align*}
Now we compute the determinant of the matrix via its eigenvalues, the matrix has three distinct eigenvalues: 
\begin{itemize}
	\item $\lambda_{1} = a-b$,  $(n-1)$-fold degenerate
	\item $ \lambda_{2} = \dfrac{1}{2} (x - \sqrt{y}) $
	\item $ \lambda_{3} = \dfrac{1}{2} (x + \sqrt{y}) $,\\ 
	      with
		  $x = a + (n-1) b + e$ \\
		  $y = (a-e)^{2} + (a + (n-1)b)^{2} -a^{2} -2(n-1)b e + 4ncd$.
\end{itemize}
Thus, one obtain 
\begin{align}
\nonumber \text{ln}[\det (\beta M \Sigma + I)] =& (n-1)\text{ln}(a-b) + \text{ln}[\dfrac{1}{4}(x^{2}- y)]  \\
\nonumber = & (n-1)\text{ln}(a-b) + \text{ln}[ae + (n-1)be  \\
            & - ncd]
\end{align}
now substituting $c$ and $e$ then using the identity Eq.~(\ref{Eq:repZ}) yields
\small
\begin{align}
\nonumber &\dfrac{\partial}{\partial n}(2 \alpha G_{r})\mid_{n=0} ~  =  \\
\nonumber &\alpha \left[  \dfrac{- a \beta (w-t)+ b \beta (w-t)+ b + \beta(w-t) \overbrace{\beta (u-v)}^{(a-b)-1}}{a-b} + \text{ln}(a-b) \right] \\
 &= \alpha\left[ \text{ln}(a-b)+ \dfrac{\beta (v-w) - \beta (w-t)}{a-b}\right]  
\end{align}
\normalsize
Finally, we insert the expressions of $a,b$ and $t$  one obtain the energetic term
\begin{align}
\dfrac{\partial}{\partial n}\left[ 2 \alpha G_{r}\right] _{n=0} &= \alpha\left[  \dfrac{\beta(v - 2w + 1/2 - 1/2\pi)}{1+ \beta (u-v)} + \text{ln}[1+ \beta (u-v)] \right]  .
\end{align}
Proceeding to the calculations of the entropic term, the $ [n+1]K $- square matrix $\mathcal{C}$ has the block form 
\begin{align}
\mathcal{C} = \begin{pmatrix}
		Q^{nK \times nK}  & R^{nK \times K} \\
		R^{T}  & T^{K \times K}
		 \end{pmatrix} \; ,
\end{align}
since we have assumed an orthonormal teacher vectors, the teacher-teacher overlaps block $T^{K \times K} $ is just a $K \times K$ unit matrix. While using the ansatz of the order parameters Eq.~(\ref{Eq:ansat_rep}), the student-student overlaps $Q^{nK \times nK} $ and the student-teacher overlaps $R^{nK \times K} $ blocks takes the form
\begin{align}
Q^{nK \times nK} = \begin{pmatrix}
						Q_{ij}^{aa} &Q_{ij}^{ab} &\cdots &Q_{ij}^{ab} \\
						Q_{ij}^{ab} &Q_{ij}^{aa} &\cdots &Q_{ij}^{ab} \\
						\vdots &\vdots &\ddots &\vdots \\
						Q_{ij}^{ab} &Q_{ij}^{ab} &\cdots &Q_{ij}^{aa}
						\end{pmatrix},
\end{align}
with
\begin{align*}
Q_{ij}^{aa} = \begin{pmatrix}
			  1 &C &\cdots &C \\
			  C &1  &\cdots &C \\
			  \vdots &\vdots &\ddots &\vdots \\
			  C &C &\cdots &1
		      \end{pmatrix} \; , \quad 
Q_{ij}^{ab} = \begin{pmatrix}
			  q &p &\cdots &p \\
			  p &q &\cdots &p \\
			  \vdots &\vdots &\ddots &\vdots \\
			  p &p &\cdots &q
			  \end{pmatrix}	,	      
\end{align*}
and
\begin{align}
R^{nK \times K} = \begin{pmatrix}
					  R_{ij}^{a}\\
					  R_{ij}^{a}\\
					   \vdots \\
					  R_{ij}^{a}
					  \end{pmatrix} ,
\end{align}
with
\begin{align*}
R_{ij}^{a} = \begin{pmatrix}
				  R &S &\cdots &S \\
				  S &R &\cdots &S \\
				  \vdots &\vdots &\ddots &\vdots \\
				  S &S &\cdots &R
				  \end{pmatrix}.
\end{align*}
Similar to the calculations of the energetic term, we compute $(\det \mathcal{C})$ through its eigenvalues but first we apply Schur complement for the determinant of block matrices which simplify the calculations of the eigenvalues. The Schur complement states that
\begin{align}
\det  \begin{pmatrix}
		 A^{n \times n}  & B^{n \times m} \\
		 C^{m \times n}  & D^{m \times m}
		 \end{pmatrix} =  \det (D) \, \det (A - B D^{-1} C)
		 \; ,
\end{align}
hence, we obtain
\begin{align}
\nonumber \det (\mathcal{C}) &= \det (T) \, \det (Q -RT^{-1}R^{T})\\
\nonumber			   &= \det (Q -RR^{T})
\end{align}
Diagonalization of the $nK \times nK$ matrix yields four distinct eigenvalues 
\begin{itemize}
 	\item $ \lambda_{1} = \tilde{a} + (K-1) \tilde{b} + (n-1) \tilde{c} + (n-1)(K-1) \tilde{d}  $. 
 	\item $ \lambda_{2} = \tilde{a} + (K-1) \tilde{b} - \tilde{c} - (K-1) \tilde{d} $, 
 	      $n-1$-fold degenerate. 
 	\item $ \lambda_{3} = \tilde{a} - \tilde{b} + (n-1) \tilde{c} - (n-1) \tilde{d} $,
 		  $(K-1)$-fold degenerate.
    \item $ \lambda_{4} = \tilde{a} - \tilde{b} - \tilde{c} + \tilde{d} $, 
          $(n-1)(K-1)$-fold degenerate. 
\end{itemize}
Here, we have defined the abbreviations 
\begin{align*}
\tilde{a} &= 1- R^{2} - (K-1)S^{2}, \quad
\tilde{b} = C - 2RS - (K-2)S^{2} \\
\tilde{c} &= q- R^{2} - (K-1) S^{2}, \quad
\tilde{d} = p - 2RS - (K-2) S^{2}
\end{align*}
Thus, the entropic term of the free energy is computed by
\small
\begin{align}
\dfrac{\partial}{\partial n}(\dfrac{2}{K} G_{s})\mid_{n=0}  = \dfrac{1}{K} \dfrac{\partial}{\partial n} \left[ \underbrace{\text{ln}(\lambda_{1})}_{I} + \underbrace{\text{ln}(\lambda_{2})}_{II} + \underbrace{\text{ln}(\lambda_{3})}_{III} + \underbrace{\text{ln}(\lambda_{4})}_{IV} \right]_{n=0} .
\end{align}
\normalsize
Substituting the explicit expressions of $\tilde{a}, \tilde{b}, \tilde{c},$ and $\tilde{d}$ then rewriting the results in terms of $\Delta, \delta, \hat{p}, \hat{S}$, we compute the terms I to IV as 
\begin{enumerate}[label=\Roman*.]
	\item \begin{align*}
		 \dfrac{1}{K} \dfrac{\partial}{\partial n} \text{ln}(\lambda_{1})\vert_{n=0} = \dfrac{\tilde{c} + (K-1)\tilde{d} }{\tilde{a}+      
		    (K-1)\tilde{b} - \tilde{c} - (K-1) \tilde{d} }
		  \end{align*}
    the numerator yields
           \begin{align*}	    
		     &= (q-R^{2} - (K-1)S^{2}) + (K-1) (p- 2RS - (K-1)S^{2})\\
		     &= \delta + \hat{p} - (R^{2} + S^{2} - 2RS) - 2K RS - K^{2} S^{2} + 2K S^{2}\\
		     &= \delta + \hat{p} - \Delta^{2} - \underbrace{2KRS - K^{2} S^{2} + 2 K S^{2}}_{
		                                             2K(\Delta + S)S - \hat{S}^{2}+ 2KS^{2}} \\
		     &= \delta + \hat{p} - \Delta^{2} - 2 \Delta \hat{S} - \hat{S}^{2} \\
		     &= \delta + \hat{p} - (\Delta + \hat{S})^{2}                         
		   \end{align*}
    using similar calculations the denominator term yields
	      \begin{align*}
	        &= K(1 + \hat{C} - (\Delta + \hat{S})^{2} - \delta - \hat{p}+
	              (\Delta + \hat{S})^{2}) \\
	        &= \underbrace{K(1 + \hat{C} - \delta - \hat{p} )}_{\tilde{C}}
	      \end{align*}
	  hence one obtain \\
	   \begin{align}
      \dfrac{1}{K} \dfrac{\partial}{\partial n} \text{ln}(\lambda_{1})\vert_{n=0} = 
       \dfrac{\delta + \hat{p} - (\Delta + \hat{S})^{2}}{\tilde{C}} 
       \end{align}
    \item This term is sub-leading of order $\mathcal{O}(1/K)$, hence it can be neglected in the large $K$ limit.
    \item  
     \begin{align*}
    \dfrac{1}{K} \dfrac{\partial}{\partial n} \text{ln}(\lambda_{3})\vert_{n=0} &= \dfrac{K-1}{K} \dfrac{\partial}{\partial n} \text{ln}(\tilde{a} - \tilde{b} 
         + (n-1) \tilde{c} - (n-1)\tilde{d})\vert_{n=0}\\
      &= \dfrac{K-1}{K}  \dfrac{\tilde{c} - \tilde{d}}{\tilde{a} - \tilde{b}- \tilde{c}
           + \tilde{d}} \\
      &= \dfrac{K-1}{K}  \dfrac{q-p- R^{2} + 2RS - S^{2}}{1-C -q+p} \\
      &= (1- \dfrac{1}{K})  \dfrac{\delta - \Delta^{2}}{1-\delta - \dfrac{\hat{C}}{K-1}}
     \end{align*}
     For large $K$ one obtain \\
     \begin{align}
      \dfrac{1}{K} \dfrac{\partial}{\partial n} \text{ln}(\lambda_{3})\vert_{n=0}  - \dfrac{\delta - \Delta^{2}}{\delta - 1}
     \end{align}
     \item  
     \begin{align*}
     \dfrac{1}{K} \dfrac{\partial}{\partial n} \text{ln}(\lambda_{4})\vert_{n=0} 
     &= \dfrac{K-1}{K}  \text{ln}(\tilde{a} - \tilde{b} - \tilde{C} + \tilde{d})\\
     &= (1- \dfrac{1}{K})  \text{ln} (1 - \delta - \dfrac{\hat{C}}{K-1} )
     \end{align*}
      Which in the large $K$ limit yields \\
      \begin{align}
          \dfrac{1}{K} \dfrac{\partial}{\partial n} \text{ln}(\lambda_{4})\vert_{n=0}  = \text{ln}(1-\delta) .
      \end{align}       
\end{enumerate} 
Collecting all the terms yields the entropic term 
\begin{align}
\nonumber \dfrac{\partial}{\partial n} \left[- \dfrac{2}{K} G_{s} \right]_{n=0} =& \dfrac{\delta - \Delta^{2}}{\delta - 1} - \text{ln}(1-\delta) \\ 
& - \dfrac{\delta + \hat{p} - (\Delta + \hat{S})^{2}}{\tilde{C}} + \mathcal{O}(\dfrac{1}{K})
\end{align}

\section{ The Quenched Free Energy saddle point calculations }
Here we compute the saddle point equations and solutions of the free energy Eq.~(\ref{Eq:free_rep}), using that $\hat{C} = \tilde{C}/K + \delta + \hat{p} -1 \approx \delta + \hat{p} -1$ one can eliminate $\hat{C}$ accordingly. Next we compute the derivatives 
\begin{align*}
\dfrac{\partial f}{\partial \hat{p}} = \dfrac{\partial f}{\partial \hat{S}} = \dfrac{\partial f}{\partial \tilde{C}} = \dfrac{\partial f}{\partial \delta} = \dfrac{\partial f}{\partial \Delta} = 0 ,
\end{align*}
one obtains
\begin{widetext}
\begin{align}
 \dfrac{\alpha \beta /4}{1+ \beta \left( \dfrac{1}{4} - \dfrac{\sqrt{1- \delta^{2}}}{2 \pi} - \dfrac{\delta \arcsin[\delta]}{2\pi}\right) } - \dfrac{1}{\tilde{C}} &= 0 \label{Eq:saddle1} \\
 \dfrac{2(\Delta + \hat{S})}{\tilde{C}} - \dfrac{\alpha \beta /2}{1+ \beta \left( \dfrac{1}{4} - \dfrac{\sqrt{1- \delta^{2}}}{2 \pi} - \dfrac{\delta \arcsin[\delta]}{2\pi}\right) } &= 0  \label{Eq:saddle2} \\
 \dfrac{\hat{p} + \delta - (\Delta+\hat{S})^{2}}{\tilde{C}^{2}} &= 0 \label{Eq:saddle3} \\
 -\dfrac{\alpha \beta \left( \dfrac{1}{4} + \dfrac{\arcsin[\Delta]}{2\pi} \right) }{1+ \beta \left( \dfrac{1}{4} - \dfrac{\sqrt{1- \delta^{2}}}{2 \pi} - \dfrac{\delta \arcsin[\delta]}{2\pi}\right)} - \dfrac{2 \Delta}{\delta -1} + \dfrac{2(\Delta + \hat{S})}{\tilde{C}} &= 0  \label{Eq:saddle4} 
\end{align}
\small
\begin{align}
\nonumber & 
 \alpha \left[  \dfrac{\beta/4}{1+ \beta \left( \dfrac{1}{4} - \dfrac{\sqrt{1- \delta^{2}}}{2 \pi} - \dfrac{\delta \arcsin[\delta]}{2\pi}\right)}  
+ \dfrac{\beta^{2} \arcsin[\delta] \left(  \dfrac{1}{2} + \dfrac{\hat{p}}{4} + \dfrac{\delta}{4}+ \dfrac{\sqrt{1- \delta^{2}}}{2 \pi} + \dfrac{\delta \arcsin[\delta]}{2\pi} -  \dfrac{\hat{S}}{2} - \dfrac{\Delta}{2} - \dfrac{\sqrt{1- \Delta^{2}}}{\pi} - \dfrac{\Delta \arcsin[\Delta]}{\pi}  \right)  }{2\pi \left( 1+ \beta \left( \dfrac{1}{4} - \dfrac{\sqrt{1- \delta^{2}}}{2 \pi} - \dfrac{\delta \arcsin[\delta]}{2\pi}\right)\right)^{2} }
\right. \\ 
& \left. + \dfrac{\beta/4}{1+ \beta \left( \dfrac{1}{4} - \dfrac{\sqrt{1- \delta^{2}}}{2 \pi} - \dfrac{\delta \arcsin[\delta]}{2\pi}\right)}  
\right] - \dfrac{\delta - \Delta^{2}}{(\delta-1)^{2}} - \dfrac{1}{\tilde{C}} = 0  \label{Eq:saddle5}
\end{align}
\normalsize
\end{widetext}
From Eq.~(\ref{Eq:saddle1}) one obtain
\begin{align}
\dfrac{1}{\tilde{C}} = \dfrac{\alpha \beta /4}{1+ \beta \left( \dfrac{1}{4} - \dfrac{\sqrt{1- \delta^{2}}}{2 \pi} - \dfrac{\delta \arcsin[\delta]}{2\pi}\right) } ,
\label{Eq:saddle1_1} 
\end{align}
substitute $1/\tilde{C}$ in Eq.~(\ref{Eq:saddle2}) yields $\hat{S} = 1- \Delta$. Now substituting $\hat{S}$ in Eq.~(\ref{Eq:saddle3}) one finds $\hat{p} = 1- \delta$, note that for these solutions to exist one need to assume that $\tilde{C}$ is of $\mathcal{O}(1)$. Consequently in the limit $K \rightarrow \infty$, one should assume $\hat{C} \rightarrow 0$ such that $\tilde{C}$ is of order one. Finally substituting the solutions of $\hat{S},\hat{p}$ and Eq.~(\ref{Eq:saddle1_1}) into Eq.~(\ref{Eq:saddle4}) and Eq.~(\ref{Eq:saddle5}) yields
\begin{widetext}
\begin{align}
& \dfrac{\Delta}{\delta -1} + \dfrac{\alpha \beta \left( \dfrac{\arcsin[\Delta]}{2\pi} \right) }{1+ \beta \left( \dfrac{1}{4} - \dfrac{\sqrt{1- \delta^{2}}}{2 \pi} - \dfrac{\delta \arcsin[\delta]}{2\pi}\right)} = 0 , \label{Eq:saddle_I} \\
& \alpha \left[ \dfrac{\beta^{2} \arcsin[\delta] \left(  \dfrac{1}{4}+ \dfrac{\sqrt{1- \delta^{2}}}{2 \pi} + \dfrac{\delta \arcsin[\delta]}{2\pi} - \dfrac{\sqrt{1- \Delta^{2}}}{\pi} - \dfrac{\Delta \arcsin[\Delta]}{\pi}  \right)  }{2\pi \left( 1+ \beta \left( \dfrac{1}{4} - \dfrac{\sqrt{1- \delta^{2}}}{2 \pi} - \dfrac{\delta \arcsin[\delta]}{2\pi}\right)\right)^{2} }
\right] - \dfrac{\delta - \Delta^{2}}{(\delta-1)^{2}} = 0 . \label{Eq:saddle_II} 
\end{align}
\end{widetext}
For finite values of $\beta$, one needs to solve these equations numerically  to find $(\Delta,\delta)$ as a function of $(\alpha,\beta)$. Which yields $\Delta=\delta = 0$ in the unspecialized phase and $\Delta(\delta) > 0$ for $\alpha > \alpha_{c}$ in the specialized regime. 
\onecolumngrid
\section*{Refrences}
\twocolumngrid     
\bibliography{biblog}

\begin{thebibliography}{49}%
\makeatletter
\providecommand \@ifxundefined [1]{%
 \@ifx{#1\undefined}
}%
\providecommand \@ifnum [1]{%
 \ifnum #1\expandafter \@firstoftwo
 \else \expandafter \@secondoftwo
 \fi
}%
\providecommand \@ifx [1]{%
 \ifx #1\expandafter \@firstoftwo
 \else \expandafter \@secondoftwo
 \fi
}%
\providecommand \natexlab [1]{#1}%
\providecommand \enquote  [1]{``#1''}%
\providecommand \bibnamefont  [1]{#1}%
\providecommand \bibfnamefont [1]{#1}%
\providecommand \citenamefont [1]{#1}%
\providecommand \href@noop [0]{\@secondoftwo}%
\providecommand \href [0]{\begingroup \@sanitize@url \@href}%
\providecommand \@href[1]{\@@startlink{#1}\@@href}%
\providecommand \@@href[1]{\endgroup#1\@@endlink}%
\providecommand \@sanitize@url [0]{\catcode `\\12\catcode `\$12\catcode `\&12\catcode `\#12\catcode `\^12\catcode `\_12\catcode `\%12\relax}%
\providecommand \@@startlink[1]{}%
\providecommand \@@endlink[0]{}%
\providecommand \url  [0]{\begingroup\@sanitize@url \@url }%
\providecommand \@url [1]{\endgroup\@href {#1}{\urlprefix }}%
\providecommand \urlprefix  [0]{URL }%
\providecommand \Eprint [0]{\href }%
\providecommand \doibase [0]{https://doi.org/}%
\providecommand \selectlanguage [0]{\@gobble}%
\providecommand \bibinfo  [0]{\@secondoftwo}%
\providecommand \bibfield  [0]{\@secondoftwo}%
\providecommand \translation [1]{[#1]}%
\providecommand \BibitemOpen [0]{}%
\providecommand \bibitemStop [0]{}%
\providecommand \bibitemNoStop [0]{.\EOS\space}%
\providecommand \EOS [0]{\spacefactor3000\relax}%
\providecommand \BibitemShut  [1]{\csname bibitem#1\endcsname}%
\let\auto@bib@innerbib\@empty
\bibitem [{\citenamefont {Mathew}\ \emph {et~al.}(2021)\citenamefont {Mathew}, \citenamefont {Amudha},\ and\ \citenamefont {Sivakumari}}]{Deepbook}%
  \BibitemOpen
  \bibfield  {author} {\bibinfo {author} {\bibfnamefont {A.}~\bibnamefont {Mathew}}, \bibinfo {author} {\bibfnamefont {P.}~\bibnamefont {Amudha}},\ and\ \bibinfo {author} {\bibfnamefont {S.}~\bibnamefont {Sivakumari}},\ }\bibfield  {title} {\bibinfo {title} {Deep learning techniques: An overview},\ }in\ \href@noop {} {\emph {\bibinfo {booktitle} {Advanced Machine Learning Technologies and Applications}}},\ \bibinfo {editor} {edited by\ \bibinfo {editor} {\bibfnamefont {A.~E.}\ \bibnamefont {Hassanien}}, \bibinfo {editor} {\bibfnamefont {R.}~\bibnamefont {Bhatnagar}},\ and\ \bibinfo {editor} {\bibfnamefont {A.}~\bibnamefont {Darwish}}}\ (\bibinfo  {publisher} {Springer Singapore},\ \bibinfo {address} {Singapore},\ \bibinfo {year} {2021})\ pp.\ \bibinfo {pages} {599--608}\BibitemShut {NoStop}%
\bibitem [{\citenamefont {Collins}\ \emph {et~al.}(2021)\citenamefont {Collins}, \citenamefont {Dennehy}, \citenamefont {Conboy},\ and\ \citenamefont {Mikalef}}]{COLLINS2021}%
  \BibitemOpen
  \bibfield  {author} {\bibinfo {author} {\bibfnamefont {C.}~\bibnamefont {Collins}}, \bibinfo {author} {\bibfnamefont {D.}~\bibnamefont {Dennehy}}, \bibinfo {author} {\bibfnamefont {K.}~\bibnamefont {Conboy}},\ and\ \bibinfo {author} {\bibfnamefont {P.}~\bibnamefont {Mikalef}},\ }\bibfield  {title} {\bibinfo {title} {Artificial intelligence in information systems research: A systematic literature review and research agenda},\ }\href {https://doi.org/https://doi.org/10.1016/j.ijinfomgt.2021.102383} {\bibfield  {journal} {\bibinfo  {journal} {International Journal of Information Management}\ }\textbf {\bibinfo {volume} {60}},\ \bibinfo {pages} {102383} (\bibinfo {year} {2021})}\BibitemShut {NoStop}%
\bibitem [{\citenamefont {Niskanen}\ \emph {et~al.}(2023)\citenamefont {Niskanen}, \citenamefont {Sipola},\ and\ \citenamefont {Väänänen}}]{niskanen2023}%
  \BibitemOpen
  \bibfield  {author} {\bibinfo {author} {\bibfnamefont {T.}~\bibnamefont {Niskanen}}, \bibinfo {author} {\bibfnamefont {T.}~\bibnamefont {Sipola}},\ and\ \bibinfo {author} {\bibfnamefont {O.}~\bibnamefont {Väänänen}},\ }\href {https://arxiv.org/abs/2305.04532} {\bibinfo {title} {Latest trends in artificial intelligence technology: A scoping review}} (\bibinfo {year} {2023}),\ \Eprint {https://arxiv.org/abs/2305.04532} {arXiv:2305.04532 [cs.LG]} \BibitemShut {NoStop}%
\bibitem [{\citenamefont {LeCun}\ \emph {et~al.}(2015)\citenamefont {LeCun}, \citenamefont {Bengio},\ and\ \citenamefont {Hinton}}]{lecun2015}%
  \BibitemOpen
  \bibfield  {author} {\bibinfo {author} {\bibfnamefont {Y.}~\bibnamefont {LeCun}}, \bibinfo {author} {\bibfnamefont {Y.}~\bibnamefont {Bengio}},\ and\ \bibinfo {author} {\bibfnamefont {G.}~\bibnamefont {Hinton}},\ }\bibfield  {title} {\bibinfo {title} {Deep learning},\ }\href@noop {} {\bibfield  {journal} {\bibinfo  {journal} {Nature}\ }\textbf {\bibinfo {volume} {521}},\ \bibinfo {pages} {436} (\bibinfo {year} {2015})}\BibitemShut {NoStop}%
\bibitem [{\citenamefont {Goodfellow}\ \emph {et~al.}(2016)\citenamefont {Goodfellow}, \citenamefont {Bengio},\ and\ \citenamefont {Courville}}]{goodfellow2016}%
  \BibitemOpen
  \bibfield  {author} {\bibinfo {author} {\bibfnamefont {I.}~\bibnamefont {Goodfellow}}, \bibinfo {author} {\bibfnamefont {Y.}~\bibnamefont {Bengio}},\ and\ \bibinfo {author} {\bibfnamefont {A.}~\bibnamefont {Courville}},\ }\href@noop {} {\emph {\bibinfo {title} {Deep Learning}}}\ (\bibinfo  {publisher} {MIT Press},\ \bibinfo {year} {2016})\BibitemShut {NoStop}%
\bibitem [{\citenamefont {Engel}\ and\ \citenamefont {Van~den Broeck}(2001)}]{engel}%
  \BibitemOpen
  \bibfield  {author} {\bibinfo {author} {\bibfnamefont {A.}~\bibnamefont {Engel}}\ and\ \bibinfo {author} {\bibfnamefont {C.}~\bibnamefont {Van~den Broeck}},\ }\href@noop {} {\emph {\bibinfo {title} {Statistical Mechanics of Learning}}}\ (\bibinfo  {publisher} {Cambridge University Press},\ \bibinfo {year} {2001})\BibitemShut {NoStop}%
\bibitem [{\citenamefont {Biehl}(2022)}]{book_shallow}%
  \BibitemOpen
  \bibfield  {author} {\bibinfo {author} {\bibfnamefont {M.}~\bibnamefont {Biehl}},\ }\href {https://doi.org/10.21827/648c59c1a467e} {\emph {\bibinfo {title} {The Shallow and the Deep: A biased introduction to neural networks and old school machine learning}}}\ (\bibinfo  {publisher} {University of Groningen},\ \bibinfo {year} {2022})\BibitemShut {NoStop}%
\bibitem [{\citenamefont {Nishimori}(2001{\natexlab{a}})}]{nishimori2001}%
  \BibitemOpen
  \bibfield  {author} {\bibinfo {author} {\bibfnamefont {H.}~\bibnamefont {Nishimori}},\ }\href@noop {} {\emph {\bibinfo {title} {Statistical Physics of Spin Glasses and Information Processing: An Introduction}}}\ (\bibinfo  {publisher} {Oxford University Press},\ \bibinfo {year} {2001})\BibitemShut {NoStop}%
\bibitem [{\citenamefont {Watkin}\ \emph {et~al.}(1993)\citenamefont {Watkin}, \citenamefont {Rau},\ and\ \citenamefont {Biehl}}]{Watkin}%
  \BibitemOpen
  \bibfield  {author} {\bibinfo {author} {\bibfnamefont {T.~L.~H.}\ \bibnamefont {Watkin}}, \bibinfo {author} {\bibfnamefont {A.}~\bibnamefont {Rau}},\ and\ \bibinfo {author} {\bibfnamefont {M.}~\bibnamefont {Biehl}},\ }\bibfield  {title} {\bibinfo {title} {The statistical mechanics of learning a rule},\ }\href {https://doi.org/10.1103/RevModPhys.65.499} {\bibfield  {journal} {\bibinfo  {journal} {Rev. Mod. Phys.}\ }\textbf {\bibinfo {volume} {65}},\ \bibinfo {pages} {499} (\bibinfo {year} {1993})}\BibitemShut {NoStop}%
\bibitem [{\citenamefont {Advani}\ \emph {et~al.}(2013)\citenamefont {Advani}, \citenamefont {Lahiri},\ and\ \citenamefont {Ganguli}}]{Advani}%
  \BibitemOpen
  \bibfield  {author} {\bibinfo {author} {\bibfnamefont {M.}~\bibnamefont {Advani}}, \bibinfo {author} {\bibfnamefont {S.}~\bibnamefont {Lahiri}},\ and\ \bibinfo {author} {\bibfnamefont {S.}~\bibnamefont {Ganguli}},\ }\bibfield  {title} {\bibinfo {title} {Statistical mechanics of complex neural systems and high dimensional data},\ }\href {https://doi.org/10.1088/1742-5468/2013/03/P03014} {\bibfield  {journal} {\bibinfo  {journal} {Journal of Statistical Mechanics: Theory and Experiment}\ }\textbf {\bibinfo {volume} {2013}},\ \bibinfo {pages} {P03014} (\bibinfo {year} {2013})}\BibitemShut {NoStop}%
\bibitem [{\citenamefont {Bahri}\ \emph {et~al.}(2020)\citenamefont {Bahri}, \citenamefont {Kadmon}, \citenamefont {Pennington}, \citenamefont {Schoenholz}, \citenamefont {Sohl-Dickstein},\ and\ \citenamefont {Ganguli}}]{phase_Bahri}%
  \BibitemOpen
  \bibfield  {author} {\bibinfo {author} {\bibfnamefont {Y.}~\bibnamefont {Bahri}}, \bibinfo {author} {\bibfnamefont {J.}~\bibnamefont {Kadmon}}, \bibinfo {author} {\bibfnamefont {J.}~\bibnamefont {Pennington}}, \bibinfo {author} {\bibfnamefont {S.~S.}\ \bibnamefont {Schoenholz}}, \bibinfo {author} {\bibfnamefont {J.}~\bibnamefont {Sohl-Dickstein}},\ and\ \bibinfo {author} {\bibfnamefont {S.}~\bibnamefont {Ganguli}},\ }\bibfield  {title} {\bibinfo {title} {Statistical mechanics of deep learning},\ }\href {https://doi.org/10.1146/annurev-conmatphys-031119-050745} {\bibfield  {journal} {\bibinfo  {journal} {Annual Review of Condensed Matter Physics}\ }\textbf {\bibinfo {volume} {11}},\ \bibinfo {pages} {501} (\bibinfo {year} {2020})},\ \Eprint {https://arxiv.org/abs/https://doi.org/10.1146/annurev-conmatphys-031119-050745} {https://doi.org/10.1146/annurev-conmatphys-031119-050745} \BibitemShut {NoStop}%
\bibitem [{\citenamefont {Zdeborov{\'a}}\ and\ \citenamefont {Krzakala}(2016)}]{Zdeborova2016InferenceReview}%
  \BibitemOpen
  \bibfield  {author} {\bibinfo {author} {\bibfnamefont {L.}~\bibnamefont {Zdeborov{\'a}}}\ and\ \bibinfo {author} {\bibfnamefont {F.}~\bibnamefont {Krzakala}},\ }\bibfield  {title} {\bibinfo {title} {Statistical physics of inference: Thresholds and algorithms},\ }\href {https://doi.org/10.1080/00018732.2016.1211393} {\bibfield  {journal} {\bibinfo  {journal} {Advances in Physics}\ }\textbf {\bibinfo {volume} {65}},\ \bibinfo {pages} {453} (\bibinfo {year} {2016})}\BibitemShut {NoStop}%
\bibitem [{\citenamefont {Seung}\ \emph {et~al.}(1992)\citenamefont {Seung}, \citenamefont {Sompolinsky},\ and\ \citenamefont {Tishby}}]{seung1992}%
  \BibitemOpen
  \bibfield  {author} {\bibinfo {author} {\bibfnamefont {H.~S.}\ \bibnamefont {Seung}}, \bibinfo {author} {\bibfnamefont {H.}~\bibnamefont {Sompolinsky}},\ and\ \bibinfo {author} {\bibfnamefont {N.}~\bibnamefont {Tishby}},\ }\bibfield  {title} {\bibinfo {title} {Statistical mechanics of learning from examples},\ }\href@noop {} {\bibfield  {journal} {\bibinfo  {journal} {Physical Review A}\ }\textbf {\bibinfo {volume} {45}},\ \bibinfo {pages} {6056} (\bibinfo {year} {1992})}\BibitemShut {NoStop}%
\bibitem [{\citenamefont {Opper}(1996)}]{opper1996}%
  \BibitemOpen
  \bibfield  {author} {\bibinfo {author} {\bibfnamefont {M.}~\bibnamefont {Opper}},\ }\bibfield  {title} {\bibinfo {title} {Statistical mechanics of generalization},\ }in\ \href@noop {} {\emph {\bibinfo {booktitle} {The Handbook of Brain Theory and Neural Networks}}},\ \bibinfo {editor} {edited by\ \bibinfo {editor} {\bibfnamefont {M.~A.}\ \bibnamefont {Arbib}}}\ (\bibinfo  {publisher} {MIT Press},\ \bibinfo {year} {1996})\ pp.\ \bibinfo {pages} {922--925}\BibitemShut {NoStop}%
\bibitem [{\citenamefont {Schwarze}\ and\ \citenamefont {Hertz}(1992)}]{H.Schwarze_1992}%
  \BibitemOpen
  \bibfield  {author} {\bibinfo {author} {\bibfnamefont {H.}~\bibnamefont {Schwarze}}\ and\ \bibinfo {author} {\bibfnamefont {J.}~\bibnamefont {Hertz}},\ }\bibfield  {title} {\bibinfo {title} {Generalization in a large committee machine},\ }\href {https://doi.org/10.1209/0295-5075/20/4/015} {\bibfield  {journal} {\bibinfo  {journal} {Europhysics Letters}\ }\textbf {\bibinfo {volume} {20}},\ \bibinfo {pages} {375} (\bibinfo {year} {1992})}\BibitemShut {NoStop}%
\bibitem [{\citenamefont {Schwarze}\ and\ \citenamefont {Hertz}(1993{\natexlab{a}})}]{H.Schwarze_1993}%
  \BibitemOpen
  \bibfield  {author} {\bibinfo {author} {\bibfnamefont {H.}~\bibnamefont {Schwarze}}\ and\ \bibinfo {author} {\bibfnamefont {J.}~\bibnamefont {Hertz}},\ }\bibfield  {title} {\bibinfo {title} {Generalization in fully connected committee machines},\ }\href {https://doi.org/10.1209/0295-5075/21/7/012} {\bibfield  {journal} {\bibinfo  {journal} {Europhysics Letters}\ }\textbf {\bibinfo {volume} {21}},\ \bibinfo {pages} {785} (\bibinfo {year} {1993}{\natexlab{a}})}\BibitemShut {NoStop}%
\bibitem [{\citenamefont {Schwarze}\ and\ \citenamefont {Hertz}(1993{\natexlab{b}})}]{HSchwarze2_1993}%
  \BibitemOpen
  \bibfield  {author} {\bibinfo {author} {\bibfnamefont {H.}~\bibnamefont {Schwarze}}\ and\ \bibinfo {author} {\bibfnamefont {J.}~\bibnamefont {Hertz}},\ }\bibfield  {title} {\bibinfo {title} {Learning from examples in fully connected committee machines},\ }\href {https://doi.org/10.1088/0305-4470/26/19/024} {\bibfield  {journal} {\bibinfo  {journal} {Journal of Physics A: Mathematical and General}\ }\textbf {\bibinfo {volume} {26}},\ \bibinfo {pages} {4919} (\bibinfo {year} {1993}{\natexlab{b}})}\BibitemShut {NoStop}%
\bibitem [{\citenamefont {Schwarze}(1993)}]{HSchwarze_1993}%
  \BibitemOpen
  \bibfield  {author} {\bibinfo {author} {\bibfnamefont {H.}~\bibnamefont {Schwarze}},\ }\bibfield  {title} {\bibinfo {title} {Learning a rule in a multilayer neural network},\ }\href {https://doi.org/10.1088/0305-4470/26/21/017} {\bibfield  {journal} {\bibinfo  {journal} {Journal of Physics A: Mathematical and General}\ }\textbf {\bibinfo {volume} {26}},\ \bibinfo {pages} {5781} (\bibinfo {year} {1993})}\BibitemShut {NoStop}%
\bibitem [{\citenamefont {Ahr}\ \emph {et~al.}(1999)\citenamefont {Ahr}, \citenamefont {Biehl},\ and\ \citenamefont {Urbanczik}}]{Ahr_1999}%
  \BibitemOpen
  \bibfield  {author} {\bibinfo {author} {\bibfnamefont {M.}~\bibnamefont {Ahr}}, \bibinfo {author} {\bibfnamefont {M.}~\bibnamefont {Biehl}},\ and\ \bibinfo {author} {\bibfnamefont {R.}~\bibnamefont {Urbanczik}},\ }\bibfield  {title} {\bibinfo {title} {Statistical physics and practical training of soft-committee machines},\ }\href {https://doi.org/10.1007/s100510050889} {\bibfield  {journal} {\bibinfo  {journal} {The European Physical Journal B}\ }\textbf {\bibinfo {volume} {10}},\ \bibinfo {pages} {583} (\bibinfo {year} {1999})}\BibitemShut {NoStop}%
\bibitem [{\citenamefont {Aubin}\ \emph {et~al.}(2019)\citenamefont {Aubin}, \citenamefont {Maillard}, \citenamefont {Barbier}, \citenamefont {Krzakala}, \citenamefont {Macris},\ and\ \citenamefont {Zdeborov{\'a}}}]{Aubin2019CommitteeMachine}%
  \BibitemOpen
  \bibfield  {author} {\bibinfo {author} {\bibfnamefont {B.}~\bibnamefont {Aubin}}, \bibinfo {author} {\bibfnamefont {A.}~\bibnamefont {Maillard}}, \bibinfo {author} {\bibfnamefont {J.}~\bibnamefont {Barbier}}, \bibinfo {author} {\bibfnamefont {F.}~\bibnamefont {Krzakala}}, \bibinfo {author} {\bibfnamefont {N.}~\bibnamefont {Macris}},\ and\ \bibinfo {author} {\bibfnamefont {L.}~\bibnamefont {Zdeborov{\'a}}},\ }\bibfield  {title} {\bibinfo {title} {The committee machine: computational to statistical gaps in learning a two-layers neural network},\ }\href {https://doi.org/10.1088/1742-5468/ab43d2} {\bibfield  {journal} {\bibinfo  {journal} {Journal of Statistical Mechanics: Theory and Experiment}\ }\textbf {\bibinfo {volume} {2019}},\ \bibinfo {pages} {124023} (\bibinfo {year} {2019})}\BibitemShut {NoStop}%
\bibitem [{\citenamefont {Goldt}\ \emph {et~al.}(2020)\citenamefont {Goldt}, \citenamefont {Advani}, \citenamefont {Saxe}, \citenamefont {Krzakala},\ and\ \citenamefont {Zdeborov{\'a}}}]{Goldt2020DynamicsSGDTeacherStudent}%
  \BibitemOpen
  \bibfield  {author} {\bibinfo {author} {\bibfnamefont {S.}~\bibnamefont {Goldt}}, \bibinfo {author} {\bibfnamefont {M.~S.}\ \bibnamefont {Advani}}, \bibinfo {author} {\bibfnamefont {A.~M.}\ \bibnamefont {Saxe}}, \bibinfo {author} {\bibfnamefont {F.}~\bibnamefont {Krzakala}},\ and\ \bibinfo {author} {\bibfnamefont {L.}~\bibnamefont {Zdeborov{\'a}}},\ }\bibfield  {title} {\bibinfo {title} {Dynamics of stochastic gradient descent for two-layer neural networks in the teacher-student setup},\ }\href {https://doi.org/10.1088/1742-5468/abc61e} {\bibfield  {journal} {\bibinfo  {journal} {Journal of Statistical Mechanics: Theory and Experiment}\ }\textbf {\bibinfo {volume} {2020}},\ \bibinfo {pages} {124010} (\bibinfo {year} {2020})}\BibitemShut {NoStop}%
\bibitem [{\citenamefont {Goldt}\ \emph {et~al.}(2022)\citenamefont {Goldt}, \citenamefont {Loureiro}, \citenamefont {Reeves}, \citenamefont {Krzakala}, \citenamefont {M{\'e}zard},\ and\ \citenamefont {Zdeborov{\'a}}}]{Goldt2022GaussianEquivalence}%
  \BibitemOpen
  \bibfield  {author} {\bibinfo {author} {\bibfnamefont {S.}~\bibnamefont {Goldt}}, \bibinfo {author} {\bibfnamefont {B.}~\bibnamefont {Loureiro}}, \bibinfo {author} {\bibfnamefont {G.}~\bibnamefont {Reeves}}, \bibinfo {author} {\bibfnamefont {F.}~\bibnamefont {Krzakala}}, \bibinfo {author} {\bibfnamefont {M.}~\bibnamefont {M{\'e}zard}},\ and\ \bibinfo {author} {\bibfnamefont {L.}~\bibnamefont {Zdeborov{\'a}}},\ }\bibfield  {title} {\bibinfo {title} {The gaussian equivalence of generative models for learning with shallow neural networks},\ }in\ \href {https://proceedings.mlr.press/v145/goldt22a.html} {\emph {\bibinfo {booktitle} {Proceedings of the 2nd Mathematical and Scientific Machine Learning Conference}}},\ \bibinfo {series} {Proceedings of Machine Learning Research}, Vol.\ \bibinfo {volume} {145},\ \bibinfo {editor} {edited by\ \bibinfo {editor} {\bibfnamefont {J.}~\bibnamefont {Bruna}}, \bibinfo {editor} {\bibfnamefont {J.}~\bibnamefont {Hesthaven}},\ and\ \bibinfo {editor} {\bibfnamefont
  {L.}~\bibnamefont {Zdeborov{\'a}}}}\ (\bibinfo  {publisher} {PMLR},\ \bibinfo {year} {2022})\ pp.\ \bibinfo {pages} {426--471}\BibitemShut {NoStop}%
\bibitem [{\citenamefont {Nair}\ and\ \citenamefont {Hinton}(2010)}]{nair2010}%
  \BibitemOpen
  \bibfield  {author} {\bibinfo {author} {\bibfnamefont {V.}~\bibnamefont {Nair}}\ and\ \bibinfo {author} {\bibfnamefont {G.~E.}\ \bibnamefont {Hinton}},\ }\bibfield  {title} {\bibinfo {title} {Rectified linear units improve restricted boltzmann machines},\ }in\ \href@noop {} {\emph {\bibinfo {booktitle} {ICML 2010}}}\ (\bibinfo {year} {2010})\ pp.\ \bibinfo {pages} {807--814}\BibitemShut {NoStop}%
\bibitem [{\citenamefont {Glorot}\ \emph {et~al.}(2011)\citenamefont {Glorot}, \citenamefont {Bordes},\ and\ \citenamefont {Bengio}}]{glorot2011}%
  \BibitemOpen
  \bibfield  {author} {\bibinfo {author} {\bibfnamefont {X.}~\bibnamefont {Glorot}}, \bibinfo {author} {\bibfnamefont {A.}~\bibnamefont {Bordes}},\ and\ \bibinfo {author} {\bibfnamefont {Y.}~\bibnamefont {Bengio}},\ }\bibfield  {title} {\bibinfo {title} {Deep sparse rectifier neural networks},\ }in\ \href@noop {} {\emph {\bibinfo {booktitle} {Proceedings of the Fourteenth International Conference on Artificial Intelligence and Statistics}}}\ (\bibinfo {organization} {JMLR Workshop and Conference Proceedings},\ \bibinfo {year} {2011})\ pp.\ \bibinfo {pages} {315--323}\BibitemShut {NoStop}%
\bibitem [{\citenamefont {Zeiler}\ \emph {et~al.}(2013)\citenamefont {Zeiler}, \citenamefont {Ranzato}, \citenamefont {Monga}, \citenamefont {Mao}, \citenamefont {Yang}, \citenamefont {Le}, \citenamefont {Nguyen}, \citenamefont {Senior}, \citenamefont {Vanhoucke}, \citenamefont {Dean},\ and\ \citenamefont {Hinton}}]{relu_speech}%
  \BibitemOpen
  \bibfield  {author} {\bibinfo {author} {\bibfnamefont {M.}~\bibnamefont {Zeiler}}, \bibinfo {author} {\bibfnamefont {M.}~\bibnamefont {Ranzato}}, \bibinfo {author} {\bibfnamefont {R.}~\bibnamefont {Monga}}, \bibinfo {author} {\bibfnamefont {M.}~\bibnamefont {Mao}}, \bibinfo {author} {\bibfnamefont {K.}~\bibnamefont {Yang}}, \bibinfo {author} {\bibfnamefont {Q.}~\bibnamefont {Le}}, \bibinfo {author} {\bibfnamefont {P.}~\bibnamefont {Nguyen}}, \bibinfo {author} {\bibfnamefont {A.}~\bibnamefont {Senior}}, \bibinfo {author} {\bibfnamefont {V.}~\bibnamefont {Vanhoucke}}, \bibinfo {author} {\bibfnamefont {J.}~\bibnamefont {Dean}},\ and\ \bibinfo {author} {\bibfnamefont {G.}~\bibnamefont {Hinton}},\ }\bibfield  {title} {\bibinfo {title} {On rectified linear units for speech processing},\ }in\ \href@noop {} {\emph {\bibinfo {booktitle} {38th International Conference on Acoustics, Speech and Signal Processing (ICASSP)}}}\ (\bibinfo {address} {Vancouver},\ \bibinfo {year} {2013})\BibitemShut {NoStop}%
\bibitem [{\citenamefont {Xu}\ \emph {et~al.}(2015)\citenamefont {Xu}, \citenamefont {Wang}, \citenamefont {Chen},\ and\ \citenamefont {Li}}]{relu_empirical}%
  \BibitemOpen
  \bibfield  {author} {\bibinfo {author} {\bibfnamefont {B.}~\bibnamefont {Xu}}, \bibinfo {author} {\bibfnamefont {N.}~\bibnamefont {Wang}}, \bibinfo {author} {\bibfnamefont {T.}~\bibnamefont {Chen}},\ and\ \bibinfo {author} {\bibfnamefont {M.}~\bibnamefont {Li}},\ }\href@noop {} {\bibinfo {title} {Empirical evaluation of rectified activations in convolutional network}} (\bibinfo {year} {2015}),\ \Eprint {https://arxiv.org/abs/1505.00853} {arXiv:1505.00853 [cs.LG]} \BibitemShut {NoStop}%
\bibitem [{\citenamefont {Oostwal}\ \emph {et~al.}(2021)\citenamefont {Oostwal}, \citenamefont {Straat},\ and\ \citenamefont {Biehl}}]{Oostwal}%
  \BibitemOpen
  \bibfield  {author} {\bibinfo {author} {\bibfnamefont {E.}~\bibnamefont {Oostwal}}, \bibinfo {author} {\bibfnamefont {M.}~\bibnamefont {Straat}},\ and\ \bibinfo {author} {\bibfnamefont {M.}~\bibnamefont {Biehl}},\ }\bibfield  {title} {\bibinfo {title} {Hidden unit specialization in layered neural networks: Relu vs. sigmoidal activation},\ }\href {https://doi.org/10.1016/j.physa.2020.125517} {\bibfield  {journal} {\bibinfo  {journal} {Physica A: Statistical Mechanics and its Applications}\ }\textbf {\bibinfo {volume} {564}},\ \bibinfo {pages} {125517} (\bibinfo {year} {2021})}\BibitemShut {NoStop}%
\bibitem [{\citenamefont {Nishiyama}\ and\ \citenamefont {Ohzeki}(2024)}]{nishiyama2024}%
  \BibitemOpen
  \bibfield  {author} {\bibinfo {author} {\bibfnamefont {S.}~\bibnamefont {Nishiyama}}\ and\ \bibinfo {author} {\bibfnamefont {M.}~\bibnamefont {Ohzeki}},\ }\href {https://arxiv.org/abs/2404.13404} {\bibinfo {title} {Solution space and storage capacity of fully connected two-layer neural networks with generic activation functions}} (\bibinfo {year} {2024}),\ \Eprint {https://arxiv.org/abs/2404.13404} {arXiv:2404.13404 [cond-mat.dis-nn]} \BibitemShut {NoStop}%
\bibitem [{\citenamefont {Citton}\ \emph {et~al.}(2025)\citenamefont {Citton}, \citenamefont {Richert},\ and\ \citenamefont {Biehl}}]{Otavio}%
  \BibitemOpen
  \bibfield  {author} {\bibinfo {author} {\bibfnamefont {O.}~\bibnamefont {Citton}}, \bibinfo {author} {\bibfnamefont {F.}~\bibnamefont {Richert}},\ and\ \bibinfo {author} {\bibfnamefont {M.}~\bibnamefont {Biehl}},\ }\bibfield  {title} {{\selectlanguage {English}\bibinfo {title} {Phase transition analysis for shallow neural networks with arbitrary activation functions}},\ }\bibfield  {journal} {\bibinfo  {journal} {Physica A: Statistical Mechanics and its Applications}\ }\textbf {\bibinfo {volume} {660}},\ \href {https://doi.org/10.1016/j.physa.2025.130356} {10.1016/j.physa.2025.130356} (\bibinfo {year} {2025}),\ \bibinfo {note} {publisher Copyright: {\textcopyright} 2025 The Authors}\BibitemShut {NoStop}%
\bibitem [{\citenamefont {Afanah}\ and\ \citenamefont {Rosenow}(2025)}]{Afanah2025UnifiedSCM}%
  \BibitemOpen
  \bibfield  {author} {\bibinfo {author} {\bibfnamefont {A.}~\bibnamefont {Afanah}}\ and\ \bibinfo {author} {\bibfnamefont {B.}~\bibnamefont {Rosenow}},\ }\href {https://doi.org/10.48550/arXiv.2512.16556} {\bibinfo {title} {Unified description of learning dynamics in the soft committee machine from finite to ultra-wide regimes}} (\bibinfo {year} {2025}),\ \Eprint {https://arxiv.org/abs/2512.16556} {arXiv:2512.16556 [cond-mat.dis-nn]} \BibitemShut {NoStop}%
\bibitem [{\citenamefont {Engel}\ and\ \citenamefont {Reimers}(2007)}]{Engel2}%
  \BibitemOpen
  \bibfield  {author} {\bibinfo {author} {\bibfnamefont {A.}~\bibnamefont {Engel}}\ and\ \bibinfo {author} {\bibfnamefont {L.}~\bibnamefont {Reimers}},\ }\bibfield  {title} {\bibinfo {title} {Reliability of replica symmetry for the generalization problem of a toy multilayer neural network},\ }\href {https://doi.org/10.1209/0295-5075/28/7/013} {\bibfield  {journal} {\bibinfo  {journal} {EPL (Europhysics Letters)}\ }\textbf {\bibinfo {volume} {28}},\ \bibinfo {pages} {531} (\bibinfo {year} {2007})}\BibitemShut {NoStop}%
\bibitem [{\citenamefont {Dotsenko}(1995)}]{spinNN}%
  \BibitemOpen
  \bibfield  {author} {\bibinfo {author} {\bibfnamefont {V.}~\bibnamefont {Dotsenko}},\ }\href {https://doi.org/10.1142/2460} {\emph {\bibinfo {title} {An Introduction to the Theory of Spin Glasses and Neural Networks}}}\ (\bibinfo  {publisher} {WORLD SCIENTIFIC},\ \bibinfo {year} {1995})\ \Eprint {https://arxiv.org/abs/https://www.worldscientific.com/doi/pdf/10.1142/2460} {https://www.worldscientific.com/doi/pdf/10.1142/2460} \BibitemShut {NoStop}%
\bibitem [{\citenamefont {M{\'e}zard}\ \emph {et~al.}(1987)\citenamefont {M{\'e}zard}, \citenamefont {Parisi},\ and\ \citenamefont {Virasoro}}]{Mezard1987}%
  \BibitemOpen
  \bibfield  {author} {\bibinfo {author} {\bibfnamefont {M.}~\bibnamefont {M{\'e}zard}}, \bibinfo {author} {\bibfnamefont {G.}~\bibnamefont {Parisi}},\ and\ \bibinfo {author} {\bibfnamefont {M.}~\bibnamefont {Virasoro}},\ }\href@noop {} {\emph {\bibinfo {title} {Spin Glass Theory and Beyond}}}\ (\bibinfo  {publisher} {World Scientific},\ \bibinfo {year} {1987})\BibitemShut {NoStop}%
\bibitem [{\citenamefont {Nishimori}(2001{\natexlab{b}})}]{Hidetoshi}%
  \BibitemOpen
  \bibfield  {author} {\bibinfo {author} {\bibfnamefont {H.}~\bibnamefont {Nishimori}},\ }\href {https://doi.org/10.1093/acprof:oso/9780198509417.001.0001} {\emph {\bibinfo {title} {Statistical Physics of Spin Glasses and Information Processing: An Introduction}}}\ (\bibinfo  {publisher} {Oxford University Press},\ \bibinfo {year} {2001})\ \Eprint {https://arxiv.org/abs/https://academic.oup.com/book/5185/book-pdf/54038185/acprof-9780198509400.pdf} {https://academic.oup.com/book/5185/book-pdf/54038185/acprof-9780198509400.pdf} \BibitemShut {NoStop}%
\bibitem [{\citenamefont {Talagrand}(2010)}]{talagrand}%
  \BibitemOpen
  \bibfield  {author} {\bibinfo {author} {\bibfnamefont {M.}~\bibnamefont {Talagrand}},\ }\href {https://books.google.de/books?id=NR-gsTq0p-UC} {\emph {\bibinfo {title} {Mean Field Models for Spin Glasses: Volume I: Basic Examples}}},\ Ergebnisse der Mathematik und ihrer Grenzgebiete. 3. Folge / A Series of Modern Surveys in Mathematics\ (\bibinfo  {publisher} {Springer Berlin Heidelberg},\ \bibinfo {year} {2010})\BibitemShut {NoStop}%
\bibitem [{\citenamefont {de~Almeida}\ and\ \citenamefont {Thouless}(1978)}]{Almeida_1978}%
  \BibitemOpen
  \bibfield  {author} {\bibinfo {author} {\bibfnamefont {J.~R.~L.}\ \bibnamefont {de~Almeida}}\ and\ \bibinfo {author} {\bibfnamefont {D.~J.}\ \bibnamefont {Thouless}},\ }\bibfield  {title} {\bibinfo {title} {Stability of the sherrington-kirkpatrick solution of a spin glass model},\ }\href {https://doi.org/10.1088/0305-4470/11/5/028} {\bibfield  {journal} {\bibinfo  {journal} {Journal of Physics A: Mathematical and General}\ }\textbf {\bibinfo {volume} {11}},\ \bibinfo {pages} {983} (\bibinfo {year} {1978})}\BibitemShut {NoStop}%
\bibitem [{\citenamefont {Castellani}\ and\ \citenamefont {Cavagna}(2005)}]{Castellani_2005}%
  \BibitemOpen
  \bibfield  {author} {\bibinfo {author} {\bibfnamefont {T.}~\bibnamefont {Castellani}}\ and\ \bibinfo {author} {\bibfnamefont {A.}~\bibnamefont {Cavagna}},\ }\bibfield  {title} {\bibinfo {title} {Spin-glass theory for pedestrians},\ }\href {https://doi.org/10.1088/1742-5468/2005/05/P05012} {\bibfield  {journal} {\bibinfo  {journal} {Journal of Statistical Mechanics: Theory and Experiment}\ }\textbf {\bibinfo {volume} {2005}},\ \bibinfo {pages} {P05012} (\bibinfo {year} {2005})}\BibitemShut {NoStop}%
\bibitem [{\citenamefont {Malzahn}\ and\ \citenamefont {Engel}(1999)}]{Malzahn_1999}%
  \BibitemOpen
  \bibfield  {author} {\bibinfo {author} {\bibfnamefont {D.}~\bibnamefont {Malzahn}}\ and\ \bibinfo {author} {\bibfnamefont {A.}~\bibnamefont {Engel}},\ }\bibfield  {title} {\bibinfo {title} {Correlations between hidden units in multilayer neural networks and replica symmetry breaking},\ }\href {https://doi.org/10.1103/physreve.60.2097} {\bibfield  {journal} {\bibinfo  {journal} {Physical Review E}\ }\textbf {\bibinfo {volume} {60}},\ \bibinfo {pages} {2097–2104} (\bibinfo {year} {1999})}\BibitemShut {NoStop}%
\bibitem [{\citenamefont {Agliari}\ \emph {et~al.}(2020)\citenamefont {Agliari}, \citenamefont {Albanese}, \citenamefont {Barra},\ and\ \citenamefont {Ottaviani}}]{Agliari_2020}%
  \BibitemOpen
  \bibfield  {author} {\bibinfo {author} {\bibfnamefont {E.}~\bibnamefont {Agliari}}, \bibinfo {author} {\bibfnamefont {L.}~\bibnamefont {Albanese}}, \bibinfo {author} {\bibfnamefont {A.}~\bibnamefont {Barra}},\ and\ \bibinfo {author} {\bibfnamefont {G.}~\bibnamefont {Ottaviani}},\ }\bibfield  {title} {\bibinfo {title} {Replica symmetry breaking in neural networks: a few steps toward rigorous results},\ }\href {https://doi.org/10.1088/1751-8121/abaf2c} {\bibfield  {journal} {\bibinfo  {journal} {Journal of Physics A: Mathematical and Theoretical}\ }\textbf {\bibinfo {volume} {53}},\ \bibinfo {pages} {415005} (\bibinfo {year} {2020})}\BibitemShut {NoStop}%
\bibitem [{\citenamefont {Hartnett}\ \emph {et~al.}(2018)\citenamefont {Hartnett}, \citenamefont {Parker},\ and\ \citenamefont {Geist}}]{Hartnett}%
  \BibitemOpen
  \bibfield  {author} {\bibinfo {author} {\bibfnamefont {G.~S.}\ \bibnamefont {Hartnett}}, \bibinfo {author} {\bibfnamefont {E.}~\bibnamefont {Parker}},\ and\ \bibinfo {author} {\bibfnamefont {E.}~\bibnamefont {Geist}},\ }\bibfield  {title} {\bibinfo {title} {Replica symmetry breaking in bipartite spin glasses and neural networks},\ }\href {https://doi.org/10.1103/PhysRevE.98.022116} {\bibfield  {journal} {\bibinfo  {journal} {Phys. Rev. E}\ }\textbf {\bibinfo {volume} {98}},\ \bibinfo {pages} {022116} (\bibinfo {year} {2018})}\BibitemShut {NoStop}%
\bibitem [{\citenamefont {Annesi}\ \emph {et~al.}(2025)\citenamefont {Annesi}, \citenamefont {Malatesta},\ and\ \citenamefont {Zamponi}}]{Annesi2025FullRSBTwoLayer}%
  \BibitemOpen
  \bibfield  {author} {\bibinfo {author} {\bibfnamefont {B.~L.}\ \bibnamefont {Annesi}}, \bibinfo {author} {\bibfnamefont {E.~M.}\ \bibnamefont {Malatesta}},\ and\ \bibinfo {author} {\bibfnamefont {F.}~\bibnamefont {Zamponi}},\ }\bibfield  {title} {\bibinfo {title} {Exact full-rsb sat/unsat transition in infinitely wide two-layer neural networks},\ }\href {https://doi.org/10.21468/SciPostPhys.18.4.118} {\bibfield  {journal} {\bibinfo  {journal} {SciPost Physics}\ }\textbf {\bibinfo {volume} {18}},\ \bibinfo {pages} {118} (\bibinfo {year} {2025})}\BibitemShut {NoStop}%
\bibitem [{\citenamefont {Belkin}\ \emph {et~al.}(2019)\citenamefont {Belkin}, \citenamefont {Hsu}, \citenamefont {Ma},\ and\ \citenamefont {Mandal}}]{doubledecent}%
  \BibitemOpen
  \bibfield  {author} {\bibinfo {author} {\bibfnamefont {M.}~\bibnamefont {Belkin}}, \bibinfo {author} {\bibfnamefont {D.}~\bibnamefont {Hsu}}, \bibinfo {author} {\bibfnamefont {S.}~\bibnamefont {Ma}},\ and\ \bibinfo {author} {\bibfnamefont {S.}~\bibnamefont {Mandal}},\ }\bibfield  {title} {\bibinfo {title} {Reconciling modern machine learning practice and the classical bias variance trade off},\ }\href {https://doi.org/10.1073/pnas.1903070116} {\bibfield  {journal} {\bibinfo  {journal} {Proceedings of the National Academy of Sciences}\ }\textbf {\bibinfo {volume} {116}},\ \bibinfo {pages} {15849} (\bibinfo {year} {2019})},\ \Eprint {https://arxiv.org/abs/https://www.pnas.org/doi/pdf/10.1073/pnas.1903070116} {https://www.pnas.org/doi/pdf/10.1073/pnas.1903070116} \BibitemShut {NoStop}%
\bibitem [{\citenamefont {Rosen-Zvi}\ \emph {et~al.}(2001)\citenamefont {Rosen-Zvi}, \citenamefont {Engel},\ and\ \citenamefont {Kanter}}]{PhysRevLett.87.078101}%
  \BibitemOpen
  \bibfield  {author} {\bibinfo {author} {\bibfnamefont {M.}~\bibnamefont {Rosen-Zvi}}, \bibinfo {author} {\bibfnamefont {A.}~\bibnamefont {Engel}},\ and\ \bibinfo {author} {\bibfnamefont {I.}~\bibnamefont {Kanter}},\ }\bibfield  {title} {\bibinfo {title} {Multilayer neural networks with extensively many hidden units},\ }\href {https://doi.org/10.1103/PhysRevLett.87.078101} {\bibfield  {journal} {\bibinfo  {journal} {Phys. Rev. Lett.}\ }\textbf {\bibinfo {volume} {87}},\ \bibinfo {pages} {078101} (\bibinfo {year} {2001})}\BibitemShut {NoStop}%
\bibitem [{\citenamefont {Barbier}\ \emph {et~al.}(2025)\citenamefont {Barbier}, \citenamefont {Camilli}, \citenamefont {Nguyen}, \citenamefont {Pastore},\ and\ \citenamefont {Skerk}}]{Barbier25}%
  \BibitemOpen
  \bibfield  {author} {\bibinfo {author} {\bibfnamefont {J.}~\bibnamefont {Barbier}}, \bibinfo {author} {\bibfnamefont {F.}~\bibnamefont {Camilli}}, \bibinfo {author} {\bibfnamefont {M.-T.}\ \bibnamefont {Nguyen}}, \bibinfo {author} {\bibfnamefont {M.}~\bibnamefont {Pastore}},\ and\ \bibinfo {author} {\bibfnamefont {R.}~\bibnamefont {Skerk}},\ }\href {https://doi.org/10.48550/ARXIV.2510.24616} {\bibinfo {title} {Statistical physics of deep learning: Optimal learning of a multi-layer perceptron near interpolation}} (\bibinfo {year} {2025}),\ \Eprint {https://arxiv.org/abs/2510.24616} {arXiv:2510.24616} \BibitemShut {NoStop}%
\bibitem [{\citenamefont {Baldassi}\ \emph {et~al.}(2019)\citenamefont {Baldassi}, \citenamefont {Malatesta},\ and\ \citenamefont {Zecchina}}]{Baldassi2019ReLUGeometry}%
  \BibitemOpen
  \bibfield  {author} {\bibinfo {author} {\bibfnamefont {C.}~\bibnamefont {Baldassi}}, \bibinfo {author} {\bibfnamefont {E.~M.}\ \bibnamefont {Malatesta}},\ and\ \bibinfo {author} {\bibfnamefont {R.}~\bibnamefont {Zecchina}},\ }\bibfield  {title} {\bibinfo {title} {Properties of the geometry of solutions and capacity of multilayer neural networks with rectified linear unit activations},\ }\href {https://doi.org/10.1103/PhysRevLett.123.170602} {\bibfield  {journal} {\bibinfo  {journal} {Physical Review Letters}\ }\textbf {\bibinfo {volume} {123}},\ \bibinfo {pages} {170602} (\bibinfo {year} {2019})}\BibitemShut {NoStop}%
\bibitem [{\citenamefont {Steinberg}\ \emph {et~al.}(2024)\citenamefont {Steinberg}, \citenamefont {Adomaitytė}, \citenamefont {Fachechi}, \citenamefont {Mergny}, \citenamefont {Barbier},\ and\ \citenamefont {Monasson}}]{Steinberg_2024}%
  \BibitemOpen
  \bibfield  {author} {\bibinfo {author} {\bibfnamefont {J.}~\bibnamefont {Steinberg}}, \bibinfo {author} {\bibfnamefont {U.}~\bibnamefont {Adomaitytė}}, \bibinfo {author} {\bibfnamefont {A.}~\bibnamefont {Fachechi}}, \bibinfo {author} {\bibfnamefont {P.}~\bibnamefont {Mergny}}, \bibinfo {author} {\bibfnamefont {D.}~\bibnamefont {Barbier}},\ and\ \bibinfo {author} {\bibfnamefont {R.}~\bibnamefont {Monasson}},\ }\bibfield  {title} {\bibinfo {title} {Replica method for computational problems with randomness: principles and illustrations},\ }\href {https://doi.org/10.1088/1742-5468/ad292d} {\bibfield  {journal} {\bibinfo  {journal} {Journal of Statistical Mechanics: Theory and Experiment}\ }\textbf {\bibinfo {volume} {2024}},\ \bibinfo {pages} {104002} (\bibinfo {year} {2024})}\BibitemShut {NoStop}%
\bibitem [{\citenamefont {Gardner}(1988)}]{gardner1988}%
  \BibitemOpen
  \bibfield  {author} {\bibinfo {author} {\bibfnamefont {E.}~\bibnamefont {Gardner}},\ }\bibfield  {title} {\bibinfo {title} {The space of interactions in neural network models},\ }\href@noop {} {\bibfield  {journal} {\bibinfo  {journal} {Journal of Physics A: Mathematical and General}\ }\textbf {\bibinfo {volume} {21}},\ \bibinfo {pages} {257} (\bibinfo {year} {1988})}\BibitemShut {NoStop}%
\bibitem [{\citenamefont {Han}\ \emph {et~al.}(2021)\citenamefont {Han}, \citenamefont {Yao}, \citenamefont {Liu}, \citenamefont {Niu}, \citenamefont {Tsang}, \citenamefont {Kwok},\ and\ \citenamefont {Sugiyama}}]{han2021}%
  \BibitemOpen
  \bibfield  {author} {\bibinfo {author} {\bibfnamefont {B.}~\bibnamefont {Han}}, \bibinfo {author} {\bibfnamefont {Q.}~\bibnamefont {Yao}}, \bibinfo {author} {\bibfnamefont {T.}~\bibnamefont {Liu}}, \bibinfo {author} {\bibfnamefont {G.}~\bibnamefont {Niu}}, \bibinfo {author} {\bibfnamefont {I.~W.}\ \bibnamefont {Tsang}}, \bibinfo {author} {\bibfnamefont {J.~T.}\ \bibnamefont {Kwok}},\ and\ \bibinfo {author} {\bibfnamefont {M.}~\bibnamefont {Sugiyama}},\ }\href {https://arxiv.org/abs/2011.04406} {\bibinfo {title} {A survey of label-noise representation learning: Past, present and future}} (\bibinfo {year} {2021}),\ \Eprint {https://arxiv.org/abs/2011.04406} {arXiv:2011.04406 [cs.LG]} \BibitemShut {NoStop}%
\bibitem [{\citenamefont {Lehtinen}\ \emph {et~al.}(2018)\citenamefont {Lehtinen}, \citenamefont {Munkberg}, \citenamefont {Hasselgren}, \citenamefont {Laine}, \citenamefont {Karras}, \citenamefont {Aittala},\ and\ \citenamefont {Aila}}]{lehtinen2018}%
  \BibitemOpen
  \bibfield  {author} {\bibinfo {author} {\bibfnamefont {J.}~\bibnamefont {Lehtinen}}, \bibinfo {author} {\bibfnamefont {J.}~\bibnamefont {Munkberg}}, \bibinfo {author} {\bibfnamefont {J.}~\bibnamefont {Hasselgren}}, \bibinfo {author} {\bibfnamefont {S.}~\bibnamefont {Laine}}, \bibinfo {author} {\bibfnamefont {T.}~\bibnamefont {Karras}}, \bibinfo {author} {\bibfnamefont {M.}~\bibnamefont {Aittala}},\ and\ \bibinfo {author} {\bibfnamefont {T.}~\bibnamefont {Aila}},\ }\href {https://arxiv.org/abs/1803.04189} {\bibinfo {title} {Noise2noise: Learning image restoration without clean data}} (\bibinfo {year} {2018}),\ \Eprint {https://arxiv.org/abs/1803.04189} {arXiv:1803.04189 [cs.CV]} \BibitemShut {NoStop}%
\end{thebibliography}%
\nocite{*}
\end{document}